\RequirePackage{rotating}
\documentclass[twocolumn, tighten]{aastex631}
\usepackage{amsmath}
\usepackage[caption=false]{subfig}
\usepackage{xfrac}
\usepackage{CJK}

\shorttitle{Molecular Mapping of DR Tau}
\shortauthors{Huang et al.}
\begin{document}
\begin{CJK*}{UTF8}{gbsn} 
\title{Molecular Mapping of DR Tau's Protoplanetary Disk, Envelope, Outflow, and Large-Scale Spiral Arm }

\correspondingauthor{Jane Huang}
\email{jnhuang@umich.edu}
\author[0000-0001-6947-6072]{Jane Huang}
\altaffiliation{NASA Hubble Fellowship Program Sagan Fellow}
\affiliation{Department of Astronomy, University of Michigan, 323 West Hall, 1085 S. University Avenue, Ann Arbor, MI 48109, United States of America}
\author[0000-0003-4179-6394]{Edwin A. Bergin}
\affiliation{Department of Astronomy, University of Michigan, 323 West Hall, 1085 S. University Avenue, Ann Arbor, MI 48109, United States of America}
\author[0000-0001-7258-770X]{Jaehan Bae}
\affiliation{Department of Astronomy, University of Florida, Gainesville, FL 32611, United States of America}
\author[0000-0002-7695-7605]{Myriam Benisty}
\affiliation{Univ. Grenoble Alpes, CNRS, IPAG, 38000 Grenoble, France}
\author[0000-0003-2253-2270]{Sean M. Andrews} \affiliation{Center for Astrophysics \textbar\ Harvard \& Smithsonian, 60 Garden St., Cambridge, MA 02138, USA}

\begin{abstract}
DR Tau has been noted for its unusually high variability in comparison with other T Tauri stars. Although it is one of the most extensively studied pre-main sequence stars, observations with millimeter interferometry have so far been relatively limited. We present NOEMA images of $^{12}$CO, $^{13}$CO, C$^{18}$O, SO, DCO$^+$, and H$_2$CO toward DR Tau at a resolution of $\sim0.5''$ ($\sim100$ au). In addition to the protoplanetary disk, CO emission reveals an envelope, a faint asymmetric outflow, and a spiral arm with a clump. The $\sim1200$ au extent of the CO arm far exceeds that of the spiral arms previously detected in scattered light, which underlines the necessity of sensitive molecular imaging for contextualizing the disk environment. The kinematics and compact emission distribution of C$^{18}$O, SO, DCO$^+$, and H$_2$CO indicate that they originate primarily from within the Keplerian circumstellar disk. The SO emission, though, also exhibits an asymmetry that may be due to interaction with infalling material or unresolved substructure. The complex environment of DR Tau is reminiscent of those of outbursting FUor sources and some EXor sources, suggesting that DR Tau's extreme stellar activity could likewise be linked to disk instabilities promoted by large-scale infall. 
\end{abstract}


\section{Introduction} \label{sec:intro}

In the classic schema of low-mass star formation, young stellar objects (YSOs) are divided into four classes (0, I, II, and III) based on their spectral energy distributions \citep[e.g.,][]{1984ApJ...287..610L, 1987IAUS..115....1L, 1993ApJ...406..122A}. These classes are generally thought to correspond to different evolutionary stages, such that a Class 0 YSO has an envelope mass comparable to or greater than that of the protostar (and its possible disk), a Class I YSO has an envelope that is less massive than the protostar but still comparable to its disk, a Class II YSO has a disk with negligible envelope material, and a Class III YSO has negligible amounts of remaining circumstellar material \citep[e.g.,][]{1989ApJ...340..823W, 1994ApJ...420..837A, 2014prpl.conf..195D}. The correspondence between SED class, evolutionary stage, and morphology, though, is known to be imperfect \citep[e.g.,][]{2006ApJS..167..256R}. 

Planet formation models often adopt the characteristics of envelope-free Class II disks as a starting point \citep[e.g.,][]{2011ApJ...743L..16O, 2012AA...544A..32L, 2018ApJ...869L..47Z}. However, scattered light and molecular imaging have yielded identifications of a number of Class II disks that appear to be interacting either with (remnant) envelopes or ambient cloud material \citep[e.g.,][]{1999ApJ...523L.151G, 2020AA...633A..82G, 2021ApJ...908L..25G, 2022ApJ...930..171H}. The pace of these identifications has increased with the advent of instruments such as ALMA and VLT/SPHERE. Detections of gaps and rings in the millimeter continuum of some Class II disks that appear to have remnant envelope material, and even a few embedded Class I disks, offer evidence that planet formation can take place under more dynamically complex conditions than typically assumed \citep[e.g.,][]{2015ApJ...808L...3A, 2020Natur.586..228S, 2021ApJS..257...19H, 2021ApJ...909..212K}. Moreover,  \citet{2022NatAs...6..751C} recently detected a protoplanet in the disk of AB Aur, a system that appears to still be undergoing infall from a remnant envelope \citep[e.g.,][]{2012AA...547A..84T}. Simulations suggest that accretion of cloud or envelope material by the disk can influence its thermal structure, surface density profile, stability, and degree of misalignment \citep[e.g.,][]{2015ApJ...805...15B, 2019AA...628A..20D, 2022ApJ...928...92K}. These disk conditions, in turn, are expected to influence where, when, and how planets form and migrate, as well as their composition \citep[e.g.,][]{1988Icar...75..146S, 1997Sci...276.1836B, 2002ApJ...581..666K}. Hence, observations of the immediate environments of young stars are essential to establish the range of circumstances under which planet formation might proceed. 

Recent observations of DR Tau (J2000 04:47:06.215+16:58:42.81), a T Tauri star located at a distance of $192\pm1$ pc in the Taurus star-forming region \citep{2021AA...649A...1G, 2021AJ....161..147B}, have suggested that its disk is being externally perturbed. \citet{2022AA...658A..63M} detected spiral arms in scattered light images of the DR Tau protoplanetary disk and hypothesized that one of them was triggered by infalling material. Meanwhile, \citet{2022AA...660A.126S} detected non-Keplerian emission in ALMA observations of $^{13}$CO and [C I] toward DR Tau, attributing this component to an infalling stream of gas.  

DR Tau is perhaps best known for its unusual degree of stellar variability. The star has faded and brightened in B-band by several magnitudes over the course of almost a century \citep{1979AA....79L..18C}. Most notably, DR Tau brightened in B-band by about five magnitudes between 1970 to 1979, an event that \citet{1979AA....79L..18C} compared to the outbursts of FUor (also known as FU Ori) sources. DR Tau also exhibits significant short-term spectroscopic and photometric variability\textemdash on timescales of a few days, DR Tau has been observed to change by up to a couple magnitudes in B-band and by up to a factor of a few in its optical line fluxes \citep[e.g.,][]{1977AA....61..737B, 1993AA...268..192G,2001AJ....122.3335A}. DR Tau has a high stellar mass accretion rate of $4.8\times10^{-7}$ $M_\odot$ yr$^{-1}$ \citep{2019AA...632A..32M}. This high accretion level leads to significant continuum veiling, which poses a challenge for determining its spectral type \citep[e.g.,][]{1979ApJS...41..743C}. Spectral type estimates have ranged from M0 to K4 \citep[e.g.][]{1987ASSL..129..295I, 2014ApJ...786...97H, 2019AA...632A..32M, 2022AA...667A.124G}. 

DR Tau was part of the original list of outbursting EXor variables by \citet{1989ESOC...33..233H}, although it has not always been included in subsequent compilations of EXors \citep[e.g.,][]{2014prpl.conf..387A}. DR Tau is unique among the EXors listed in the \citet{1989ESOC...33..233H} catalog in that the 18-year rise time to its outburst was much longer than those of the other EXors, which were typically on the order of a couple hundred days. EXors are usually distinguished from outbursting FUor sources insofar as EXor outbursts tend to be more modest in magnitude and duration, and EXors have T Tauri-like spectra during outbursts rather than the supergiant-like spectra of FUors \citep[e.g.,][]{2014prpl.conf..387A}. Several hypotheses have been proposed to account for the outbursts of young stars, including disk instabilities driven by mass buildup through infall from envelopes or cloud material, binary interactions, and stellar flybys \citep[e.g.,][]{1992ApJ...401L..31B, 2005ApJ...633L.137V, 2010ApJ...713.1143Z, 2010MNRAS.402.1349F, 2014ApJ...795...61B, 2019AA...628A..20D}. Because these outbursts affect the disk thermal structure, they may significantly affect how planet formation proceeds by altering molecular abundances, dust properties, and snowline locations \citep[e.g.,][]{2012ApJ...744..118J, 2012ApJ...745...90B, 2016Natur.535..258C, 2018ApJ...864L..23V, 2022Natur.606..272J}. The hypothesized connection between outbursts and environmental interactions further motivates an examination of DR Tau's surroundings.  

Although DR Tau has been a popular target for observations ranging from infrared to ultraviolet wavelengths \citep[e.g.,][and references above]{1994AJ....107.2153K, 2002ApJ...566.1100A, 2008ApJ...676L..49S, 2011ApJ...733...84P, 2014ApJ...780...26B}, relatively few observations with millimeter interferometry have been reported. The millimeter continuum, which traces the distribution of large dust grains, has been imaged on several occasions \citep[e.g.,][]{2002ApJ...581..357K, 2007ApJ...659..705A, 2016AA...588A..53T, 2019ApJ...882...49L}. The millimeter continuum emission is fairly compact, with 95\% of the flux contained within a 53 au radius \citep{2019ApJ...882...49L}. Although no substructures are immediately apparent in the highest resolution image to date (tracing scales down to $\sim20$ au), modeling of the visibilities suggests the presence of gaps and rings that may be associated with planet-disk interactions \citep{2020MNRAS.495.3209J}. Other than $^{13}$CO, C$^{18}$O, and [C I] \citep{2021ApJ...908...46B, 2022AA...660A.126S}, no interferometric line observations of DR Tau have previously been published. 

The upgraded wideband capabilities of the Northern Extended Millimeter Array (NOEMA) provided an opportunity to observe a number of lines simultaneously toward DR Tau. We obtained sensitive observations of $^{12}$CO, $^{13}$CO, C$^{18}$O, SO, DCO$^+$, and H$_2$CO at a resolution of $\sim0.5''$ ($\sim100$ au) to map DR Tau's structure. The observations and data reduction are summarized in Section \ref{sec:observations}. The molecular detections are analyzed in Section \ref{sec:molecularlines}, and the implications of DR Tau's complex structures are discussed in Section \ref{sec:discussion}. The summary and conclusions are presented in Section \ref{sec:summary}.  

\section{Observations and Data Reduction}\label{sec:observations}

DR Tau was observed with the NOEMA PolyFiX correlator in dual polarization mode during program W20BE (PI: J. Huang). The correlator setup covered frequencies from 213.9-221.6 GHz and 229.4-237.2 GHz at a resolution of 2 MHz. Within these frequency ranges, we placed a series of chunks, each with a resolution of 62.5 kHz and width of 64 MHz, in order to resolve molecular lines of interest (detailed further in Section \ref{sec:molecularlines} and Appendix \ref{sec:spectroscopic}).   

The first set of observations was executed in C configuration on 2021 January 08, with baseline lengths ranging from 24 to 328 m. The second set of observations was executed in A configuration on 2021 March 03, with baseline lengths ranging from 32 to 760 m. Each configuration used eleven antennas. For each set of observations, LkH$\alpha$ 101 served as the flux calibrator, 3C 84 served as the bandpass calibrator, and 0446+112 and 0507+179 served as the phase calibrators. The on-source time was 3.0 hours in C configuration and 3.4 hours in A configuration. 

The raw data were calibrated with the NOEMA pipeline in \texttt{CLIC}, which is part of the \texttt{GILDAS} package \citep{2005sf2a.conf..721P, 2013ascl.soft05010G}. Then, the following steps were performed with the \texttt{GILDAS} \texttt{MAPPING} software. The calibrated visibilities were written out to separate $uv$-tables corresponding to the low spectral resolution, wide bandwidth data and the high spectral resolution, narrow bandwidth spectral windows. After flagging of channels with strong line emission, the wide bandwidth $uv$-tables were spectrally averaged to produce continuum $uv$-tables. For each of the four basebands, the continuum was imaged with the CLEAN algorithm and three phase self-calibration loops were performed using solution intervals of 180, 90, and 45 seconds. The self-calibration solutions were then applied to the $uv$-tables for the narrow spectral windows that fell within the corresponding basebands. Continuum subtraction was performed for each spectral window separately in the $uv$ plane by fitting a linear baseline. 

The self-calibrated, continuum-subtracted $uv$ tables were converted to measurement sets to enable imaging with the Common Astronomy Software Applications (CASA) 6.4 \citep{2022PASP..134k4501C}. Because GILDAS outputs frequencies in the rest frame of the source (i.e., the frequency that corresponds to the source systemic velocity input by the observer is the rest frequency of the line of interest), we had to manually correct the frequencies in the measurement sets so that CASA would output image cubes with the appropriate LSRK velocities. Each line was imaged with the \texttt{tclean} implementation of the multi-scale CLEAN algorithm \citep{2011AA...532A..71R}. We set the robust value to 0.5 and and the image cube channel spacing to 0.2 km s$^{-1}$. To accommodate the irregular morphology of the $^{12}$CO and $^{13}$CO $J=2-1$ emission, we employed the \texttt{auto-multithresh} algorithm \citep{2020PASP..132b4505K} to define the CLEAN masks, choosing the following parameter values after some experimentation: sidelobethreshold=2.0, noisethreshold=4.0, minbeamfrac=0.3, and negativethreshold=7.0. Initial imaging tests yielded prominent striping artifacts due to the poor $uv$ sampling of the spatially extended cloud emission, so we re-imaged these lines without baselines shorter than 20 $k\lambda$. For the other molecules, where only compact emission was detected, we used a circular CLEAN mask with a radius of $2.6''$ and included all baselines. A Gaussian $uv$ taper of $1.0''$ was used to increase sensitivity to weaker lines (i.e., lines other than $^{12}$CO, $^{13}$CO, C$^{18}$O, SO, DCO$^+$, and H$_2$CO $J_{K_aK_c} = 3_{03}-2_{02}$). After CLEANing, a primary beam correction was applied to each image cube. 

Calibrated visibilities and images can be downloaded at \url{https://zenodo.org/record/7370498#.Y7U-qezMKeB}.

\section{Results\label{sec:molecularlines}}
\subsection{Overview of Line Observations}

\begin{figure*}
\begin{center}
\includegraphics{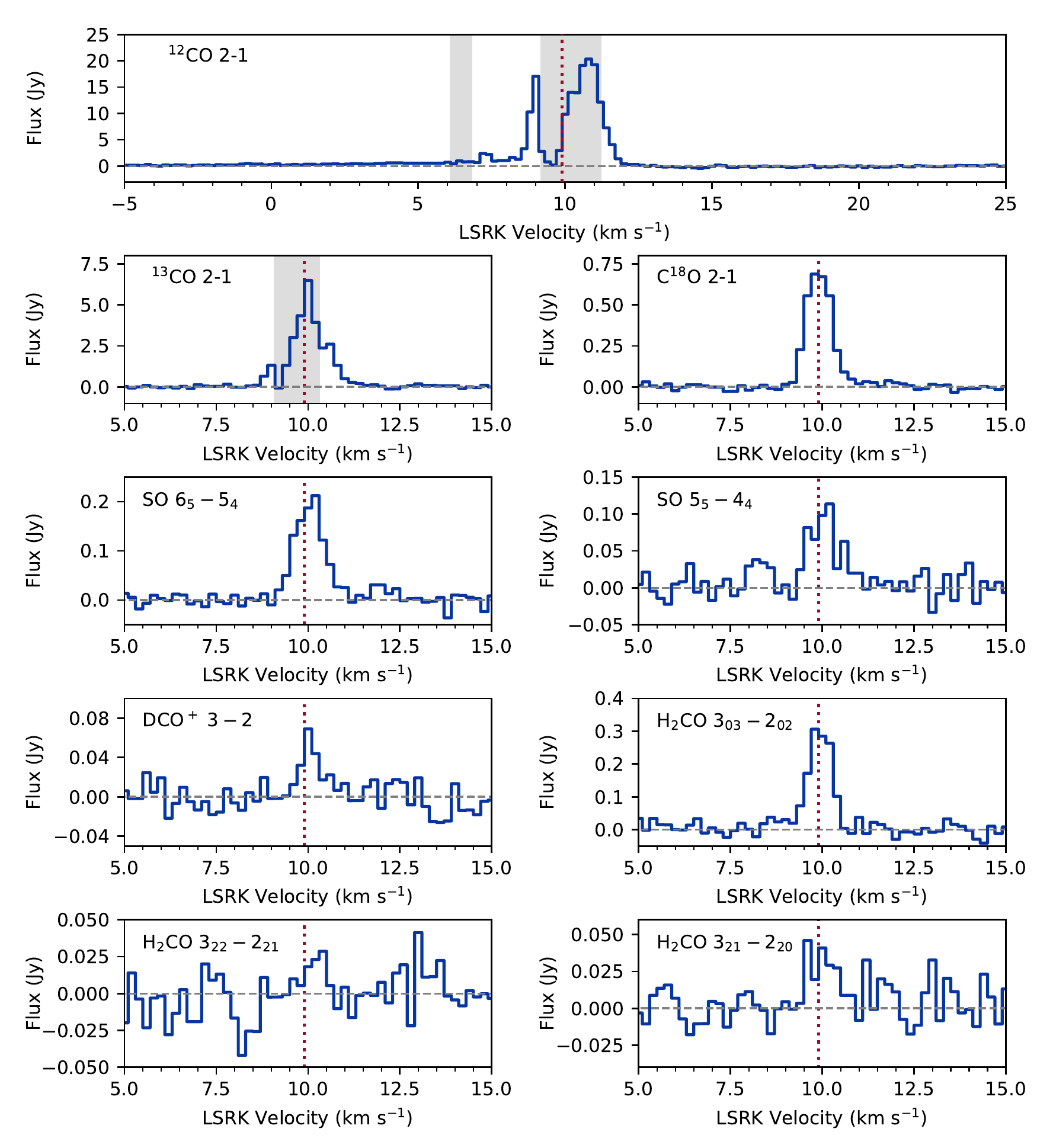}
\end{center}
\caption{Source-integrated spectra of the primary line targets toward DR Tau. The vertical red dotted line marks the systemic velocity. The gray bars denote regions where cloud contamination is apparent. \label{fig:spectraoverview}}
\end{figure*}

\begin{deluxetable*}{ccccccc}
\tablecaption{Imaging Summary for Primary Line Targets \label{tab:imageproperties}}
\tablehead{
\colhead{Transition}&\colhead{Synthesized beam}&\colhead{Per-channel RMS noise\tablenotemark{a}}&\colhead{Velocity range\tablenotemark{b}}&\colhead{Extraction Mask Diameter}&\colhead{Flux\tablenotemark{c}}\\
&(arcsec $\times$ arcsec ($^\circ$))&(mJy beam$^{-1}$)&(km s$^{-1}$)&(arcsec)&(mJy km s$^{-1}$)}
\startdata
$^{12}$CO $J=2-1$ &0.79 $\times$ 0.47 (18.2$^\circ$)&7&$[-2,17]$ &21 &$37900\pm200$\tablenotemark{d}\\
$^{13}$CO $J=2-1$ &0.84 $\times$ 0.49 (17.3$^\circ$) & 6 &$[7.6,11.4]$&21&$5730\pm70$\tablenotemark{d} \\
C$^{18}$O $J=2-1$ & 0.85 $\times$ 0.50  (17.3$^\circ$) & 6&$[9.0,10.8]$&4&$622\pm10$\\
\hline
SO $J_N = 6_5-5_4$&0.85 $\times$ 0.50  (17.2$^\circ$)&6&$[9.0,10.8]$ &3&$195\pm9$\\
SO $J_N = 5_5-4_4$ &0.86 $\times$ 0.50  (17.1$^\circ$)&6&$[9.0,10.8]$&3&$96\pm10$\\
\hline
DCO$^+$ $J=3-2$ &0.86 $\times$ 0.50  (17.1$^\circ$)&6&$[9.0,10.8]$&3&$40\pm10$\\
\hline
H$_2$CO $J_{K_aK_c}=3_{03}-2_{02}$ &0.86 $\times$ 0.50  (17.2$^\circ$)& 6&$[9.0,10.8]$&3&$248\pm11$ \\
H$_2$CO $J_{K_aK_c}=3_{22}-2_{21}$ &1.20 $\times$ 0.93  (17.1$^\circ$)&7 &$[9.0,10.8]$&3&$<30$\\
H$_2$CO $J_{K_aK_c}=3_{21}-2_{20}$ &1.20 $\times$ 0.93  (17.1$^\circ$)& 7&$[9.0,10.8]$&3&$38\pm8$\\
\enddata
\tablenotetext{a}{With channel widths of 0.2 km s$^{-1}$.}
\tablenotetext{b}{LSRK velocity range over which moment maps are produced and the flux is estimated.}
\tablenotetext{c}{The $1\sigma$ error bars do not include the systematic flux uncertainty
($\sim10\%$).}
\tablenotetext{d}{These lines are significantly affected by spatial filtering, so the statistical uncertainty does not reflect the true uncertainty in the fluxes.}
\end{deluxetable*}
The primary line targets were $^{12}$CO, $^{13}$CO, C$^{18}$O, SO, DCO$^+$, and H$_2$CO. The CO isotopologues serve as gas tracers, SO is a potential shock tracer \citep[e.g.,][]{1993MNRAS.262..915P}, and H$_2$CO and DCO$^+$ are common cold disk gas tracers \citep[e.g.,][]{2017ApJ...835..231H, 2020ApJ...890..142P}. The synthesized beam and per-channel rms (estimated from line-free channels) for the primary line targets are listed in Table \ref{tab:imageproperties}, and channel maps are presented in Appendix \ref{sec:chanmaps}. Spectra for the detected lines, which were extracted using circular masks with diameters listed in Table \ref{tab:imageproperties}, are shown in Figure \ref{fig:spectraoverview}. Since the spatial extent of $^{12}$CO and $^{13}$CO are ambiguous due to spatial filtering and cloud contamination, we used extraction masks approximately equal to the primary beam FWHM at 1.3 mm ($21''$). The mask sizes for the other lines were chosen based on the approximate radial extent of the $3\sigma$ emission in the image cubes. Fluxes were measured by integrating each spectrum within the velocity ranges listed in Table \ref{tab:imageproperties}. The velocity integration ranges for the CO isotopologues were selected based on where emission above the $3\sigma$ level is detected. For the weaker lines, the C$^{18}$O velocity integration range was adopted. The 1$\sigma$ flux uncertainties were estimated as $\Delta v \times \sqrt{N}\times \sigma_\text{spec}$, where $\Delta v$ is the channel width (in km s$^{-1}$), $N$ is the number of channels spanned by the line, and $\sigma_\text{spec}$ is the standard deviation (in Jy) measured from a signal-free portion of the spectrum (this is not to be confused with the per-channel rms value listed in Table \ref{tab:imageproperties} (in mJy beam$^{-1}$), which is calculated from the image cube). However, the statistical uncertainties do not capture the true uncertainty of the fluxes for $^{12}$CO and $^{13}$CO, which are affected by cloud contamination and spatial filtering. 

We categorize a line as detected if emission is above the $5\sigma$ level within $2''$ of DR Tau in at least one channel of the image cube and above the $3\sigma$ level in at least two adjacent channels. By these criteria, $^{12}$CO, $^{13}$CO, C$^{18}$O, SO, DCO$^+$, and H$_2$CO $3_{03}-2_{02}$ are firmly detected. While H$_2$CO $3_{21}-2_{20}$ does not meet these criteria, its integrated flux is $\gtrapprox4\sigma$ when extracted over the same velocity range and emitting region as the strong $3_{03}-2_{02}$ transition, so this line is considered to be tentatively detected. The channel maps for the $3_{22}-2_{21}$ transition (Appendix \ref{sec:chanmaps}) show $4\sigma$ emission at 10.2 kms$^{-1}$ that is cospatial with the stronger $3_{03}-2_{02}$ transition, but the velocity-integrated flux from the spectrum is $<2\sigma$. Furthermore, the peak of the spectrum occurs at a velocity well offset from the peak of the $3_{03}-2_{02}$ and $3_{21}-2_{20}$ lines. Therefore, we do not consider the $3_{22}-2_{21}$ transition to be detected. 

\begin{figure*}
\begin{center}
\includegraphics{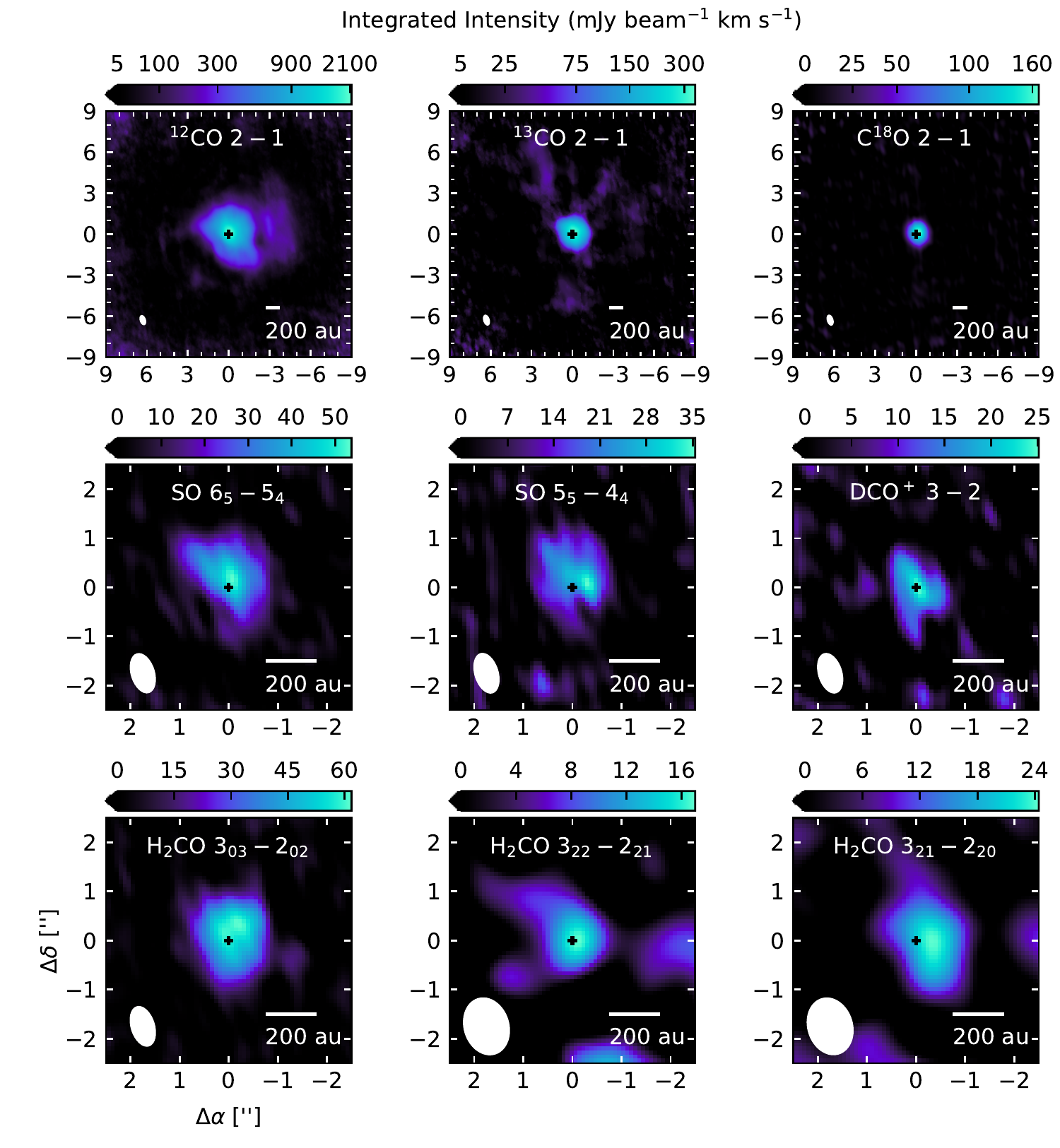}
\end{center}
\caption{Integrated intensity maps of primary line targets observed toward DR Tau. The synthesized beam is drawn in the lower left corner of each panel. Black crosses mark the disk center. The axes show offsets from the disk center in arcseconds. For the CO isotopologues, the color scale uses an arcsinh stretch to make faint extended features more visible. Note that the size scales are different between the top row and the other rows. \label{fig:mom0maps}}
\end{figure*}

\begin{figure*}
\begin{center}
\includegraphics{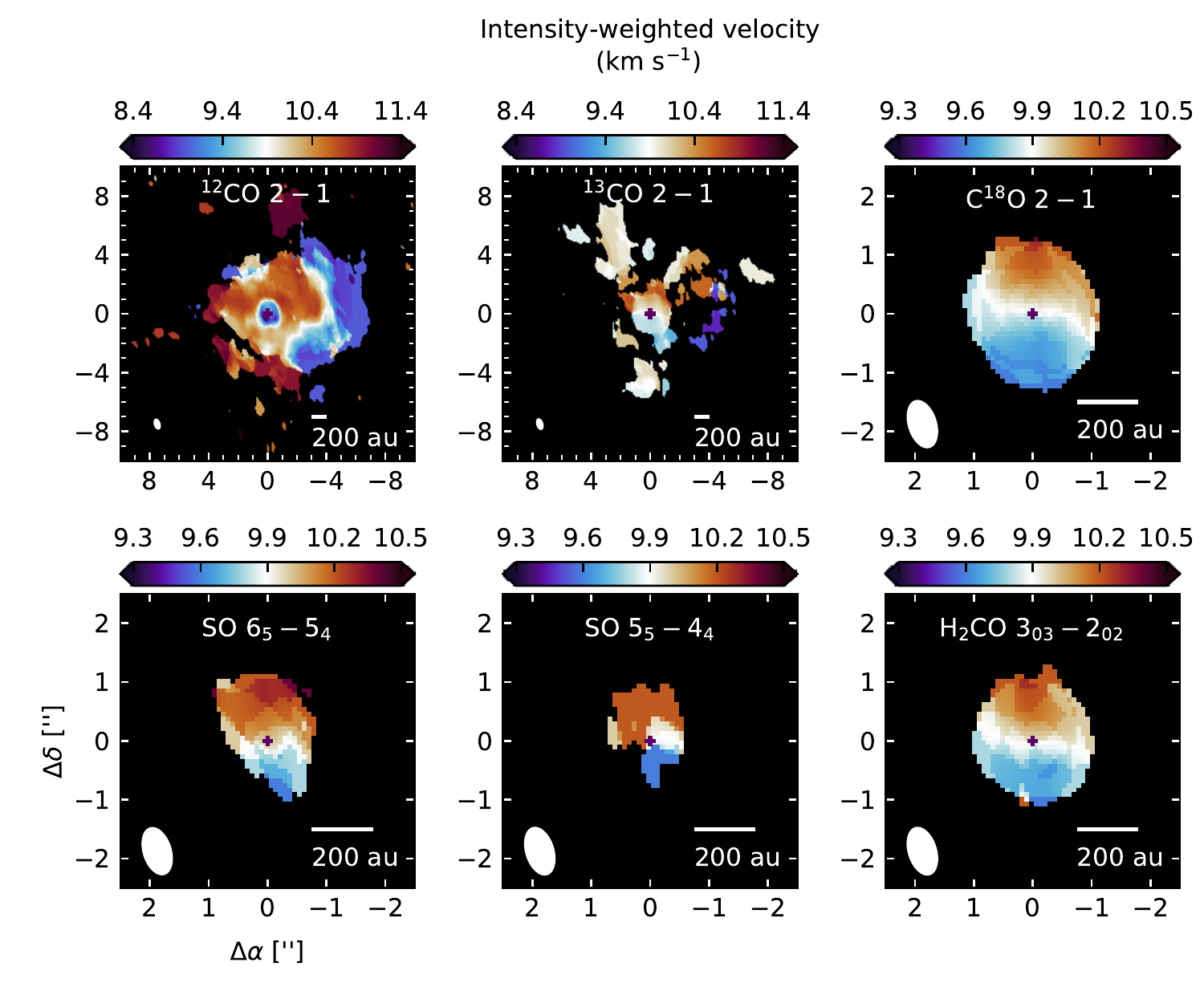}
\end{center}
\caption{Intensity-weighted velocity maps of strong lines detected toward DR Tau. The synthesized beam is drawn in the lower left corner of each panel. The purple cross denotes the disk center. The axes show offsets from the disk center in arcseconds. Note that the velocity ranges and size scales are not the same for all imags.\label{fig:mom1maps}}
\end{figure*}

Integrated intensity maps of the primary line targets are presented in Figure \ref{fig:mom0maps}, using the velocity integration ranges listed in Table \ref{tab:imageproperties}. The intensity-weighted velocity maps of the stronger lines are presented in Figure \ref{fig:mom1maps}. For $^{12}$CO and $^{13}$CO, the integrated intensity maps excluded pixels in the image cube below the $3\sigma$ level and the intensity-weighted velocity map excluded pixels below the $6\sigma$ level in order to reduce contributions from cloud contamination and artifacts from spatial filtering. For all other lines, no clipping was used for the integrated intensity maps, and a $4\sigma$ clip was adopted for the intensity-weighed velocity maps. 

A summary of the auxiliary line observations (none of which yielded a detection) is presented in Appendix \ref{sec:auxiliarylines}.

\subsection{Structures traced by CO isotopologues}
\begin{figure*}
\begin{center}
\includegraphics[scale = 0.3]{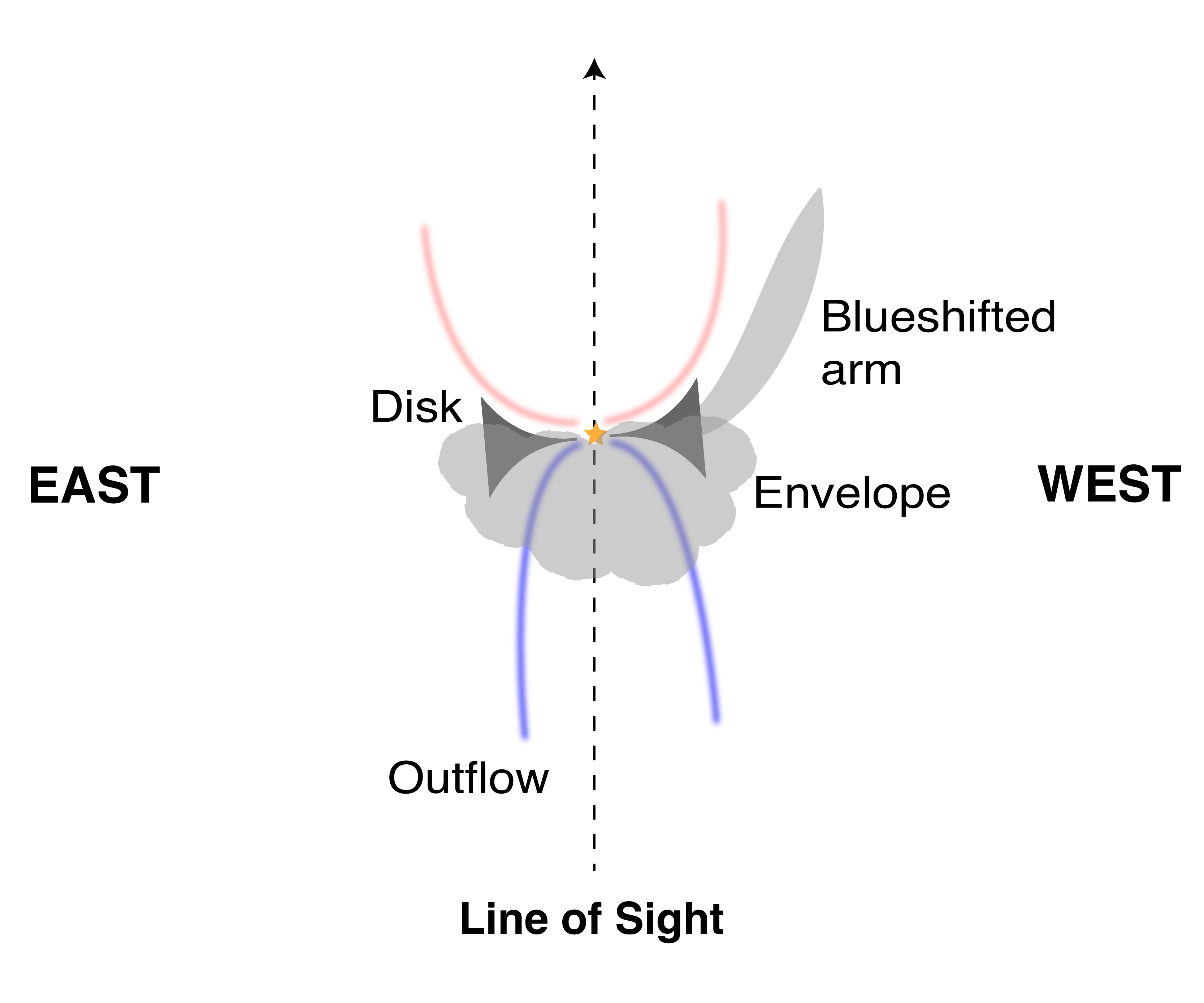}
\end{center}
\caption{A proposed cartoon schematic of the DR Tau system from an overhead perspective (i.e., perpendicular to the line of sight). The components are not drawn to scale. Note that while the disk is drawn such that the east side is tilted toward the observer in order to show that the disk is slightly inclined, the observations do not constrain which side is closer to the observer. The 3-dimensional orientations of the envelope and the arm are not known in detail, but the former is drawn in front of the disk and the latter is drawn behind the disk (from the perspective of the observer) under the assumption of infalling motion. However, the observations may also be explained by other configurations of the structures. \label{fig:DRTauschematic}}
\end{figure*}

Due to their differing optical depths, the three detected CO isotopologues reveal different components of the DR Tau system, including the circumstellar disk, an arm, an envelope, and an outflow. An overhead cartoon schematic of the system is shown in Figure \ref{fig:DRTauschematic}. We describe each component in further detail below.  

\subsubsection{The circumstellar disk}

C$^{18}$O is the least optically thick of the three detected CO isotopologues and therefore best traces the Keplerian rotation of the circumstellar disk (see Figure \ref{fig:mom1maps}). The southern side is blueshifted and the northern side is redshifted relative to the systemic velocity, which \citet{2021ApJ...908...46B} estimated to be $v_\text{sys}=9.9\substack{+0.08\\-0.09}$ km s$^{-1}$ from ALMA observations of $^{13}$CO and C$^{18}$O $J=2-1$. Signs of Keplerian rotation are visible in the inner regions of the NOEMA $^{13}$CO intensity-weighted velocity map and coincide with the bright, compact emission component in the integrated intensity map, but the disk edge is not well-defined due to the presence of extended, non-Keplerian emission. From visual inspection of the $^{13}$CO emission, we estimate that the Keplerian disk has a radial extent of $\sim300$ au, but this should only be considered a lower bound for the disk size because the abundance of $^{13}$CO is generally too low in the outer disk to recover the disk size robustly \citep[e.g.,][]{2019AA...629A..79T}. Finally, $^{12}$CO is dominated by large-scale, non-Keplerian structures. 

The NOEMA observations do not strongly constrain the disk orientation, since the C$^{18}$O emission is spanned by only a few synthesized beams. However, \citet{2019ApJ...882...49L} measured a position angle (P.A.) of $3.4\substack{+8.2\\-8.0}$ degrees east of north and an inclination angle of  $5.4\substack{+2.1\\-2.6}$ degrees from ALMA millimeter continuum observations at an angular resolution of $\sim0.1''$, which corresponds to $\sim20$ au. We adopt these values for our analysis. Although our new $^{12}$CO and $^{13}$CO observations show significant non-disk emission, \citet{2022AA...660A.126S} found that their ALMA C$^{18}$O observations could be largely reproduced by a Keplerian disk model employing the disk orientation derived from \citet{2019ApJ...882...49L}. Because the disk is nearly face-on, the C$^{18}$O spectrum only exhibits a single peak at the systemic velocity rather than the double-peak characteristic of more inclined disks.

\subsubsection{Blueshifted spiral arm\label{sec:arm}}

\begin{figure*}
\begin{center}
\includegraphics{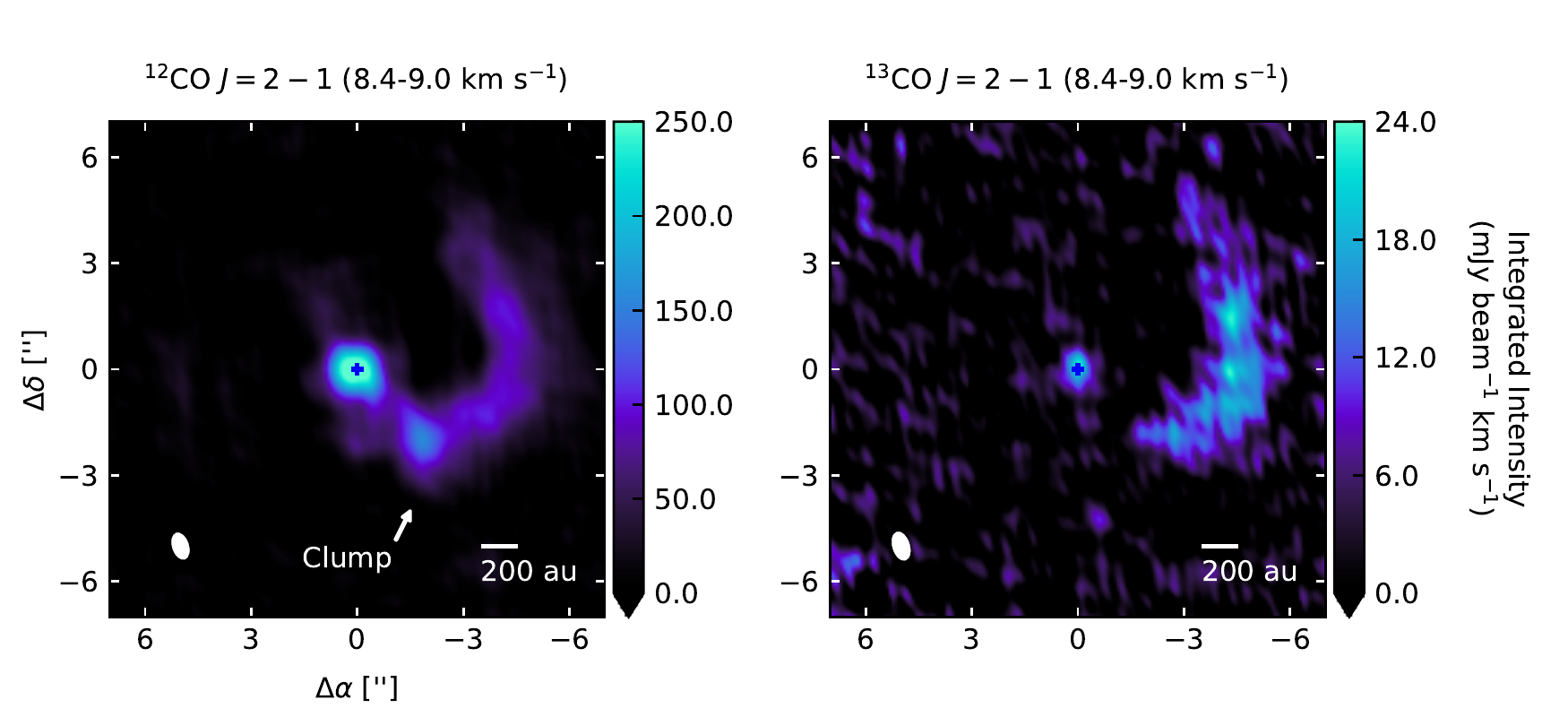}
\end{center}
\caption{Integrated intensity maps of $^{12}$CO (left) and $^{13}$CO, summed up between 8.4 and 9.0 km s$^{-1}$ to highlight DR Tau's blueshifted spiral arm. The blue cross marks the center of the disk. The synthesized beam is shown as a white ellipse in the lower left corner of each panel. The $^{12}$CO color scale is saturated in order to show the fainter arm emission more clearly. \label{fig:armmaps}}
\end{figure*}

The intensity-weighted velocity maps for $^{12}$CO and $^{13}$CO (Figure \ref{fig:mom1maps}) both show an arm that is blueshifted with respect to the systemic velocity. To isolate the emission from the CO arm, we produced integrated intensity maps between 8.4 and 9.0 km s$^{-1}$ (Figure \ref{fig:armmaps}).  The arm is connected to the south side of the disk and curves around the western side, terminating at a projected distance of $\sim1200$ au from DR Tau at a P.A. of $\sim330^\circ$. The $^{12}$CO emission also shows a clump along the arm at a projected distance of $\sim500$ au southwest from DR Tau. This arm was not detected in previously published high-resolution ALMA $^{13}$CO images of DR Tau \citep{2022AA...660A.126S}, presumably due to some combination of lack of sensitivity and spatial filtering. However, low angular resolution ALMA ACA observations of [C I] from \citet{2022AA...660A.126S} show extended blueshifted emission, which may originate from the arm traced by CO in our NOEMA observations.

\begin{figure*}
\begin{center}
\includegraphics{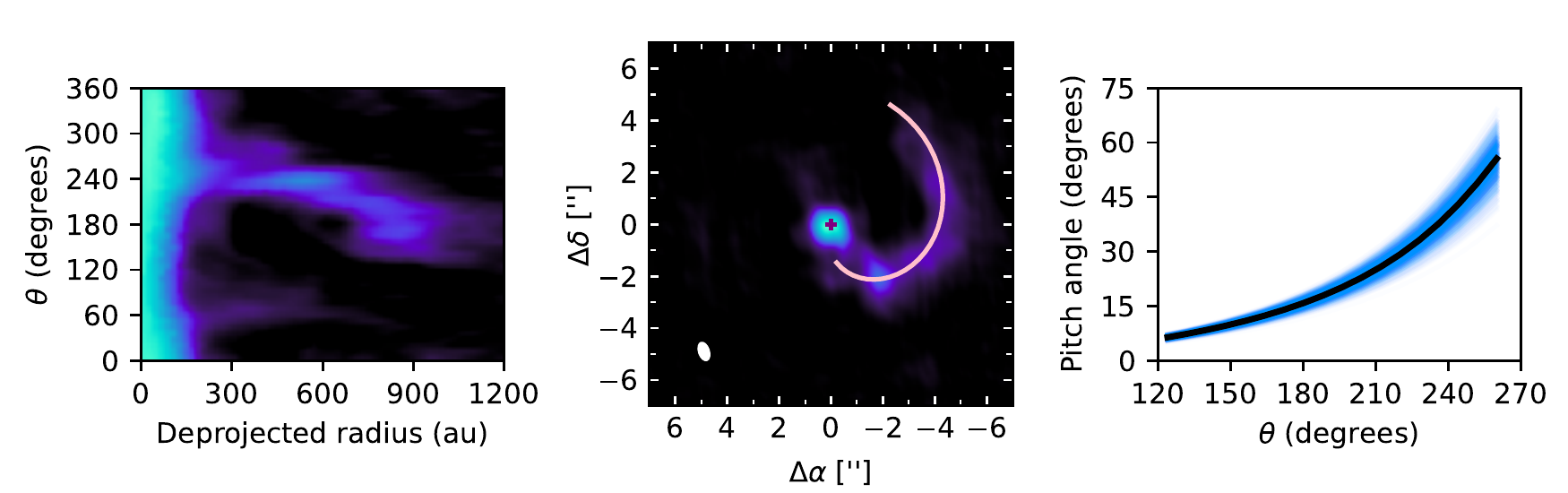}
\end{center}
\caption{Left: Integrated intensity map of the $^{12}$CO arm, replotted as a function of deprojected radius and polar angle $\theta$. Center: Integrated intensity map of the $^{12}$CO arm, overplotted with the spiral function defined by the posterior median values of the spiral parameters. Right: Pitch angle of the arm as a function of the polar angle $\theta$. The black curve corresponds to the values derived from the median of the spiral arm parameter posteriors, while the blue curves correspond to 1000 random draws from the posterior. \label{fig:pitchangle}}
\end{figure*}

In order to estimate the pitch angle of the arm, we transformed the integrated intensity map of the arm into a polar coordinate map (i.e., as a function of deprojected radius $R$ and polar angle $\theta$), assuming that the arm is in the plane of the disk (Figure \ref{fig:pitchangle}). We then measured the position of the spiral arm by searching for local radial maxima in the polar coordinate map for fixed values of $\theta$ in steps of 8$\degr$ from 124$\degr$ to 260$\degr$. The arm was modelled as an Archimedean spiral of the form $R(\theta) = a+c\theta^3$, where $\theta$ is in radians. (We found that logarithmic spirals and Archimedean spirals with smaller exponents did not fit the data well). The log-likelihood function was specified as $\log\mathcal{L}=-0.5 \sum_{n}{\left[\frac{(R_\text{data}-R_\text{model})^2}{\sigma^2}+\log(2\pi\sigma^2)\right]}$, where $\sigma$ is the standard deviation of the major axis of the synthesized beam. Uniform priors of $[0,2000]$ and $[-2000,0]$ were used for $a$ and $c$, respectively. Posteriors were explored using the affine-invariant sampler \texttt{emcee} \citep{2010CAMCS...5...65G, 2013PASP..125..306F} with 40 walkers and 1000 steps. After discarding the first 500 steps as burn-in, we computed the 50th percentile of the marginal posterior distribution to obtain a point estimate and the 16th and 84th percentiles to obtain error estimates: $a = 1060\pm30$ au and $c = -7.8\pm0.6$ au. We computed the pitch angles ($\phi =\arctan\left(\left| \frac{1}{R}\frac{dR}{d\theta}\right|
)\right)$ corresponding to the median values of $a$ and $c$, then also computed pitch angles for spiral curves defined by 1000 random draws of $(a,c)$ from the posterior. Figure \ref{fig:pitchangle} shows the median spiral plotted over the integrated intensity map and a plot of the derived pitch angles as a function of polar angle $\theta$. The pitch angles range from 6 to 56 degrees between polar angle values of 124 to 260 degrees (corresponding to deprojected radius values between 980 and 330 au). In other words, the pitch angle appears to decrease with distance from the star, although the true values may differ if the assumption that the arm is in the plane of the disk is incorrect. 

We computed the escape velocity, $v_\text{esc} = \sqrt{\frac{2GM_\ast}{r}}$, at the tip of the arm  to assess whether it is gravitationally bound to DR Tau. The dynamical mass of DR Tau has been measured to be 1.2 $M_\odot$ \citep{2021ApJ...908...46B}. Emission from the arm is detected up to $r\sim1200$ au at an LSRK velocity of 8.8 km s$^{-1}$, which is offset from the systemic velocity by 1.1 km s$^{-1}$. The corresponding escape velocity at $r=1200$ au is 1.3 km s$^{-1}$. Thus the arm appears to be compatible with being gravitationally bound to DR Tau, but not definitively so, since there may also be a transverse velocity component.

\begin{figure*}
\begin{center}
\includegraphics{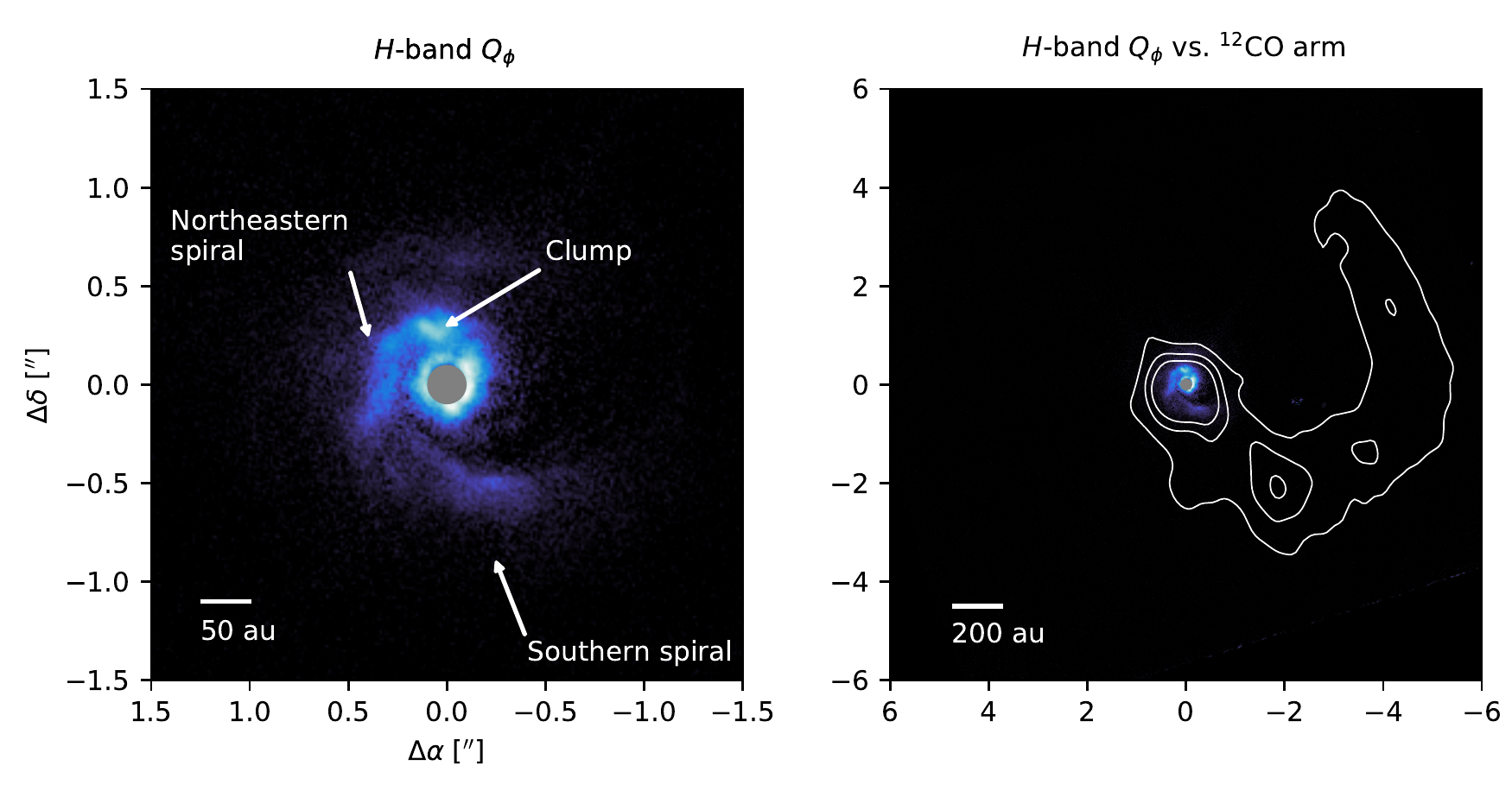}
\end{center}
\caption{A comparison between the SPHERE $H$-band $Q_\phi$ image of DR Tau from \citet{2022AA...658A..63M} and the $^{12}$CO NOEMA observations from this work. Left: $H$-band $Q_\phi$ image of DR Tau. The arrows point to the northeastern spiral, southern spiral, and clump  identified in \citet{2022AA...658A..63M}. The gray circle shows the extent of the SPHERE coronagraph. Right: A contour plot of the $^{12}$CO arm overlaid atop the $H$-band $Q_\phi$ image. Note that the size scale is different from the image on the left. The contours, drawn at 50, 100, and 150 mJy beam$^{-1}$ km s$^{-1}$, correspond to the $^{12}$CO integrated intensity map from Figure \ref{fig:armmaps}. \label{fig:SPHEREcomparison}}
\end{figure*}

\citet{2022AA...658A..63M} recently identified two spiral arms in SPHERE $H$-band $Q_\phi$ observations of DR Tau. Figure \ref{fig:SPHEREcomparison} compares the arms identified in the SPHERE image to the CO arm. The CO arm is much more extended than the scattered light arms, which are only detected up to $\sim220$ au in projection from the star. Because the NOEMA synthesized beam is comparable in scale to the SPHERE spiral arms, it is not clear whether the CO arm is an extension of one of the arms detected in scattered light or a separate structure. \citet{2022AA...658A..63M} measured pitch angles of $11\degr$ and $26\degr$ for the two scattered light arms, which are smaller than the pitch angle measured for the inner region of the CO arm. However, since the pitch angles appear to change along the arm, the differing values do not necessarily imply that they are separate structures. \citet{2022AA...658A..63M} also noted that the northeastern spiral in the SPHERE image had a clump-like feature, which they hypothesized was associated with a protoplanet embedded in a dusty envelope. While this compact feature is  well below the resolution limits of our NOEMA observations, the presence of a different clump in the $^{12}$CO arm suggests that the clumps could be intrinsic features of the arms themselves. 

\subsubsection{Envelope}
\begin{figure*}
\begin{center}
\includegraphics{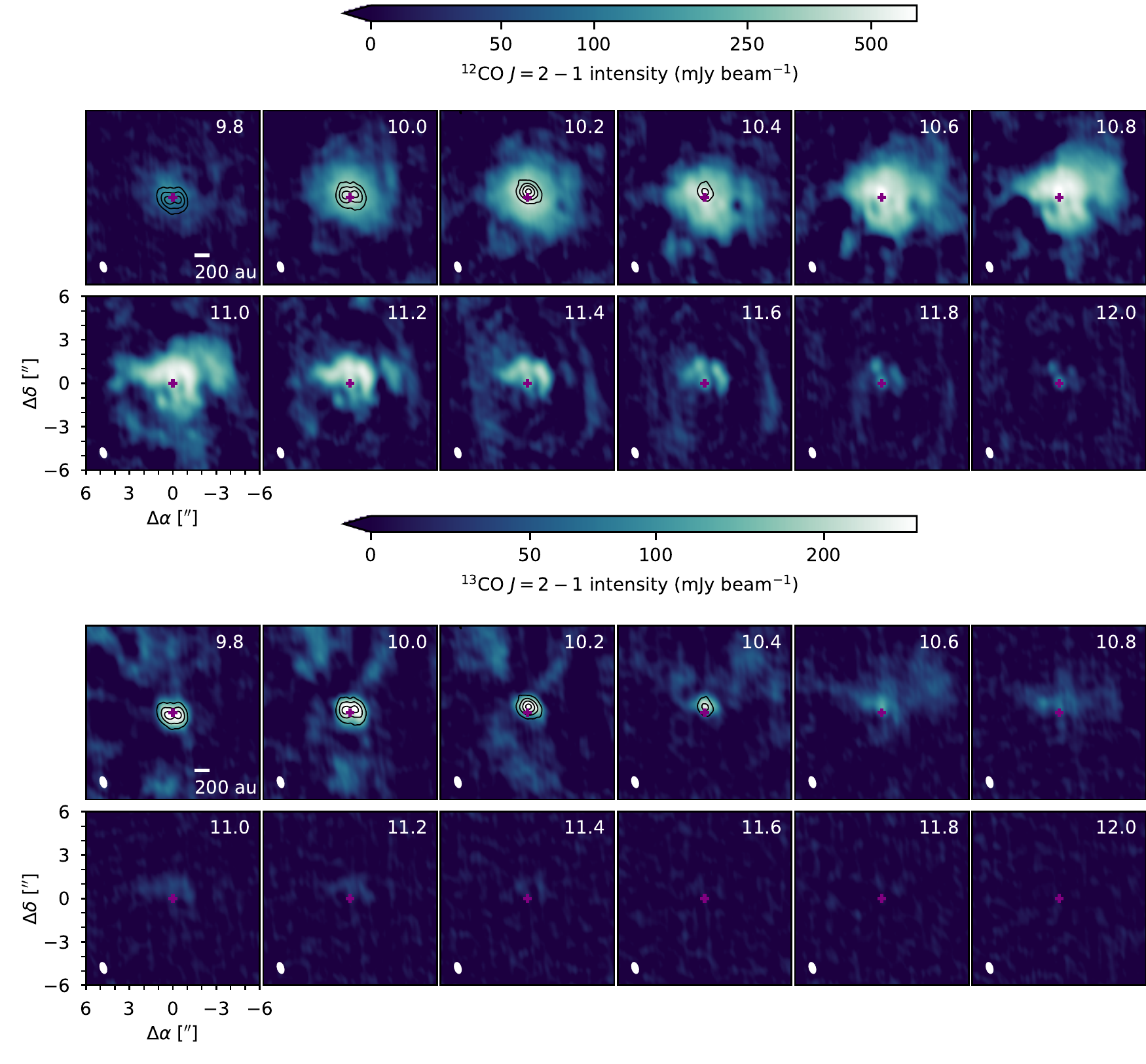}
\end{center}
\caption{Channel maps of $^{12}$CO $J=2-1$ and $^{13}$CO $J=2-1$ over the velocity range where envelope emission is present. The black contours denote the 5, 15, 25, and 35$\sigma$ contours of C$^{18}$O $J=2-1$ to serve as a visual reference for the kinematics of the Keplerian disk. (Note that because C$^{18}$O is less abundant than $^{12}$CO and $^{13}$CO, the Keplerian line wings of C$^{18}$O are not detected out to as high velocities as the other two isotopologues). \label{fig:envelope}}
\end{figure*}

DR Tau shows envelope emission in $^{12}$CO up to $\sim5''$ (1000 au) in projection from the star (Figure \ref{fig:envelope}). Envelope emission is detected between 9.8 and 12 km s$^{-1}$, i.e., mostly redshifted with respect to the systemic velocity. In most of these channels, the envelope emission is more spatially extended and brighter on the northern side. 

As with $^{12}$CO, the $^{13}$CO emission is more extended north of the star compared to south of the star for LSRK velocities above 10.4 km s$^{-1}$. In contrast to $^{12}$CO, though, the $^{13}$CO maps show features that appear more streamer-like than envelope-like. However, since this is the velocity range where cloud contamination is most significant, spatial filtering of large-scale emission may be artificially creating the appearance of streamers. \citet{2022AA...660A.126S} identified a possible infalling stream in ALMA observations of $^{13}$CO toward DR Tau, but those observations were likewise affected by spatial filtering. Observations of other lines that are bright but less susceptible to cloud contamination (e.g., species with higher critical densities like HCO$^+$ or CO transitions with higher upper energy levels) might help to clarify the nature of these apparent streamers. 

The channels where envelope emission is detected in $^{12}$CO overlap with the channels where [C I] exhibits a redshifted non-Keplerian component that \citet{2022AA...660A.126S} attributed to an infalling stream. However, since the beam FWHM of the [C I] observations is $\sim3''$, most of the emission is spatially unresolved. Given the similar velocities to the $^{12}$CO envelope, it is likely that the redshifted non-Keplerian [C I] emission also originates from the envelope. 
\subsubsection{Outflow}
\begin{figure*}
\begin{center}
\includegraphics{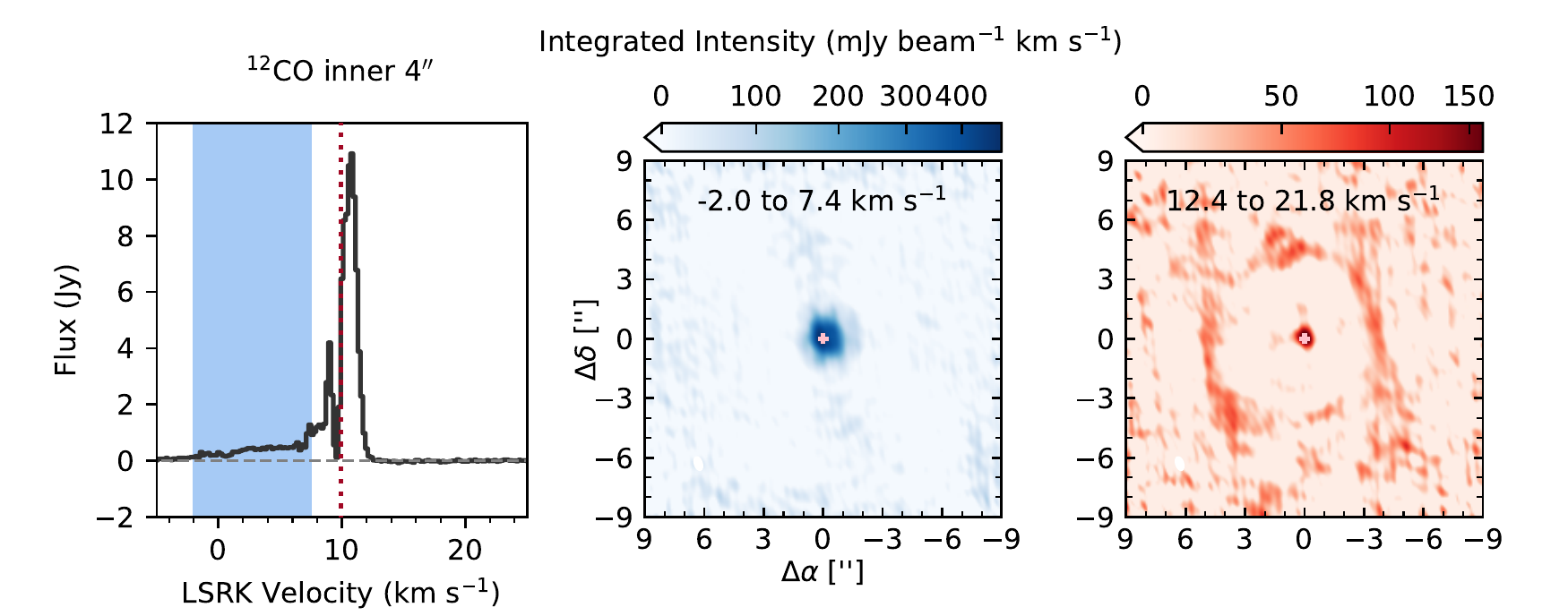}
\end{center}
\caption{Overview of DR Tau's outflow emission. Left: $^{12}$CO spectrum extracted from a circular aperture with a $4''$ diameter, showing a blueshifted outflow wing. The approximate velocity range of the blueshifted outflow is shaded in blue. The purple dotted line marks the system velocity. Middle: $^{12}$CO integrated intensity map covering velocities from $-2.0$ to 7.4 km s$^{-1}$. Compact emission from the blueshifted side of the outflow is visible. An arcsinh stretch is used on the color scale to make faint emission more readily visible. The faint vertical striping is due to the sidelobes of the point spread function.  The pink cross marks the position of the disk center. Right: $^{12}$CO integrated intensity map covering velocities from 12.4 to 21.8 km s$^{-1}$. The map shows a faint ring with a radius of $\sim4.5''$ ($\sim900$ au) and compact emission located at the stellar position. The redshifted compact emission may be from a line wing of the Keplerian disk rather than the outflow. \label{fig:outflow}}
\end{figure*}

DR Tau's $^{12}$CO spectrum (Figure \ref{fig:spectraoverview}) exhibits a faint blueshifted line wing without a corresponding redshifted line wing, suggesting the presence of an asymmetric outflow. The channel maps (Figure \ref{fig:chanmaps}) show compact emission at LSRK velocities lower than 8.4 km s$^{-1}$. To highlight this compact outflow emission more clearly, we extracted a new $^{12}$CO spectrum using a smaller circular aperture with a diameter of $4''$ (Figure \ref{fig:outflow}). Because DR Tau is nearly face-on, it is not straightforward to separate the outflow emission from the line wings of the Keplerian disk. However, the asymmetry of the line profile allows us to estimate the velocities at which outflow emission dominates by mirroring the outflow spectrum about the systemic velocity and taking the ratio of the original and mirrored spectrum. We assume that the outflow emission on the blueshifted side dominates when the ratio exceeds 10, which occurs at 7.4 km s$^{-1}$. 

While the outflow emission is weak in individual channels (Figure \ref{fig:chanmaps}), the spatial distribution of the blueshifted side can be better seen by producing an integrated intensity map between $-2.0$ and 7.4 km s$^{-1}$. The lower bound of the velocity integration range was determined by where the emission in individual channels drops below $3\sigma$. For comparison, we also produced an integrated intensity map from 12.4 to 21.8 km s$^{-1}$, corresponding to the redshifted channels at the opposing offsets from the systemic velocity. The two integrated intensity maps are presented in Figure \ref{fig:outflow}. The blueshifted map shows relatively compact emission with a radial extent of $\sim2''$ ($\sim400$ au). Although the opening angle of the outflow cannot be computed because the disk is nearly face-on, the compactness of the emission suggests that the outflow is quite collimated. The redshifted map shows emission near the stellar position, but given that the redshifted emission is fainter and much more compact than the blueshifted outflow, it seems likely that the compact redshifted emission originates from the line wing of the Keplerian disk emission. While a redshifted outflow component is not readily visible in the $^{12}$CO spectrum, the redshifted map shows a faint ring with a radius of $\sim4.5''$ ($\sim900$ au), which is significantly wider than the blueshifted outflow component.  

\subsection{SO, DCO$^+$, and H$_2$CO emission}
SO, DCO$^+$, and H$_2$CO emission all originate from a relatively compact region within 300 au of DR Tau. The SO $6_5-5_4$, SO $5_5-4_4$, and H$_2$CO $3_{03}-2_{02}$ intensity-weighted velocity maps (Figure \ref{fig:mom1maps}) show velocity gradients similar to that of C$^{18}$O, indicating that they likewise (largely) originate from the Keplerian disk. The kinematics of DCO$^+$ are not well-defined due to the low signal-to-noise ratio, but the compactness of the emission suggests that it also primarily traces the Keplerian disk. 

That said, whereas C$^{18}$O and H$_2$CO $3_{03}-2_{02}$ both exhibit relatively axisymmetric emission in the integrated intensity maps (Figure \ref{fig:mom0maps}) and line profiles that are symmetric about the systemic velocity (Figure \ref{fig:spectraoverview}), SO $6_5-5_4$ and $5_5-4_4$ are both asymmetric. Their emission is stronger on the northern (redshifted) side of the disk. In addition, their spectra both peak at an LSRK velocity of 10.2 km s$^{-1}$, which is redshifted by 0.3 km s$^{-1}$ with respect to the systemic velocity. The DCO$^+$ spectrum also appears stronger on the redshifted side, but given that its SNR is lower than that of the SO lines, more sensitive observations will be necessary to determine whether the DCO$^+$ asymmetry is genuine.

\section{Discussion \label{sec:discussion}}
\subsection{The evolutionary stage of DR Tau}
DR Tau is traditionally considered to have a Class II SED, with stellar age estimates ranging from 0.9 to 3.2 Myr  \citep{1995ApJS..101..117K, 2019AA...632A..32M, 2019ApJ...882...49L}. The presence of the envelope, if primordial, would suggest that the younger end of the age range is more likely. DR Tau's chemistry also appears to point to a younger age. Although SO is detected in DR Tau, it has otherwise rarely been detected in Class II disks, especially those hosted by T Tauri stars  \citep[e.g.,][]{2016AA...592A.124G, 2018AA...617A..28S, 2021ApJS..257...12L}. It is commonly detected, though, in younger, embedded Class 0 and I systems \citep[e.g.,][]{2014Natur.507...78S, 2020ApJ...898..131L, 2022AA...658A.104G,2022AA...659A..67M}. In addition, \citet{2022AA...660A.126S} found that gas-phase carbon is not as severely depleted in DR Tau as other Class II disks that have been observed, although it is still more depleted than Class 0/I systems.  

However, simulations have suggested that pre-main sequence stars might be able to form second-generation envelopes through interaction with cloud material, a process sometimes referred to as ``late infall'' \citep[e.g.,][]{2019AA...628A..20D, 2020AA...633A...3K}. Indeed, \citet{2022AA...658A..63M} hypothesized that DR Tau was undergoing late infall based on the detection of spiral arms in scattered light. Given the range of ages estimated for DR Tau, it is ambiguous whether infall onto DR Tau should be considered ``late.'' As noted above, DR Tau's chemistry seems to suggest that the disk is relatively young. This appearance of chemical youthfulness, though, stems from comparisons of Class 0/I sources to isolated Class II disks. Based on molecular observations of GM Aur, a Class II disk with large-scale spiral arms suggestive of ongoing late infall, \citet{2021ApJS..257...19H} speculated that accretion of cloud material could partially reset disk chemistry such that it bears greater resemblance to that of Class 0/I sources. The impact of late infall on disk chemistry will need to be examined through astrochemical modeling to determine the extent to which chemical properties can be used to sort disks by relative age. 

Given that DR Tau is commonly included in surveys of Class II disks because of its relatively large disk mass and bright line emission \citep[e.g.,][]{2011ApJ...731..130S, 2019ApJ...882...49L, 2020AJ....159..168A,2022AA...660A.126S}, an erroneous classification of its evolutionary stage may skew interpretations of disk observations. \citet{2022ApJ...930..171H} remarked that a similar problem exists for DO Tau, another commonly observed Class II disk that also shows signatures of being partially embedded. Interestingly, among the twelve single star systems that \citet{2019ApJ...882...49L} identified as having ``smooth'' disks in millimeter continuum emission, at least three of them (DR Tau, DO Tau, and Haro 6-13) exhibit evidence of an envelope in spatially resolved CO emission \citep[e.g.,][and this work]{2020AJ....159..171F, 2021AA...645A.145G, 2022ApJ...930..171H}. Most of the remaining sources (including both the ``smooth'' and structured disks) lack high quality interferometric CO observations, so it is unknown whether they might be embedded as well. Analyses of where and when disk substructures tend to emerge will require sensitive, spatially resolved molecular line observations to provide context about the evolutionary stages of the objects being studied.

\subsection{Origin of SO in DR Tau}
The detection of SO in DR Tau is notable given that SO detections have thus far been uncommon in Class II disks, which has been attributed to high gas-phase C/O ratios ($>1$) disfavoring SO production  \citep[e.g.,][]{2016AA...592A.124G, 2018AA...617A..28S, 2021ApJS..257...12L}. In two of the disks where SO has been detected, AB Aur and Oph IRS 48, the gas-phase C/O ratio has been estimated to be less than 1 \citep{2020AA...642A..32R, 2021AA...651L...6B}. Based on thermochemical modeling of [C I] and CO isotopologue emission, \citet{2022AA...660A.126S} estimated that DR Tau has a gas-phase C/O ratio of 0.47. The detection of SO toward DR Tau is thus qualitatively consistent with SO production in disks being favored in gas with C/O ratios less than 1.

DR Tau's SO emission exhibits a mild asymmetry that is not seen in C$^{18}$O. This suggests that the SO asymmetry is not merely tracing the underlying gas surface density, but could instead be due to some dynamical process locally favoring SO production. SO has been proposed to be enhanced by outflows, winds, gravitational instabilities, or accretion shocks \citep[e.g.,][]{1993MNRAS.262..915P, 2014Natur.507...78S, 2017AA...607L...6T,  2017MNRAS.472..189I}. We discuss these possibilities in turn for DR Tau.

Our NOEMA observations have shown that DR Tau has a molecular outflow, and past CO ro-vibrational spectroscopy indicates that DR Tau has a wide-angle molecular wind  \citep{2011ApJ...733...84P}. An outflow shock does not appear to be a likely major contributor to SO in DR Tau, since SO is not detected at the same high velocities as CO.  At the spatial and spectral resolution of our NOEMA observations, it is unclear whether the SO kinematics are consistent with those expected for a disk wind \citep[e.g.][]{2020MNRAS.492.5030H}, but DR Tau's bright emission makes it an excellent target for more detailed follow-up.

Chemical modeling of gravitationally unstable disks suggests that gas-phase SO can be enhanced either by spiral shocks or within warm disk fragments \citep{2011MNRAS.417.2950I, 2017MNRAS.472..189I}. While DR Tau does feature spiral structure, \citet{2022AA...658A..63M} argued that DR Tau's spiral arms are unlikely to be due to gravitational instability given that its disk-to-stellar mass ratio and stellar accretion rate are a factor of a few lower than hydrodynamical simulations suggest would be necessary to induce gravitational instabilities. Nevertheless, disk mass is notoriously difficult to measure \citep[e.g.,][and references therein]{2022arXiv220309818M}, and studies of other spiral-armed disks have often disagreed on whether they are massive enough to be gravitationally unstable \citep[e.g.,][]{2016Sci...353.1519P, 2016ApJ...832..110C, 2019MNRAS.489.3758V, 2021ApJS..257...14S}. Higher spatial resolution would help to determine if the SO asymmetry traces spiral structure and/or a disk fragment.

Accretion shocks in protoplanetary disks might occur due to cloud or envelope material being accreted by the disk or disk material being accreted by an embedded planet \citep[e.g.,][]{1974Icar...23..319B, 1993Icar..106..168B, 1999ApJ...525..330Y, 2017MNRAS.465L..64S}. Accretion streamers traced by SO have been observed in several Class I protostellar systems \citep[e.g.,][]{2022AA...658A.104G, 2022AA...667A..20A}. Given that DR Tau is now known to be partially embedded, its asymmetric SO emission may arise in a manner similar to Class I systems. \citet{2022arXiv221014820B} proposed that an SO asymmetry in the HD 100546 disk could be due to shocks from gas accreting onto an embedded planet. This likely does not account for DR Tau's SO asymmetry, since high-contrast imaging from \citet{2022AA...658A..63M} rules out the presence of a companion above several Jupiter masses at separations greater than 50 au from DR Tau.

\subsection{Origin of DR Tau's molecular spiral arm}
\citet{2022AA...658A..63M} hypothesized that the northeastern spiral arm detected in scattered light toward DR Tau is due to planet-disk interactions, while the southern arm is due to infall from cloud material. As noted in Section \ref{sec:arm}, the angular resolution of our NOEMA observations does not allow us to determine whether the CO spiral arm is an extension of either scattered light spiral arm, but the very large extent of the molecular arm suggests that it is unlikely to be generated by interactions with a bound planet. \citet{2022AA...658A..63M} placed an upper limit of several Jupiter masses on any companion farther out than 50 au from DR Tau. Given that the millimeter continuum appears smooth down to a resolution of 20 au \citep{2018ApJ...869...17L}, it is unlikely that the disk harbors massive (super-Jovian) companions within 50 au. Moreover, hydrodynamical simulations indicate that external companions exceeding several $M_\text{J}$ should create a pair of (nearly) symmetric spiral arms \citep[e.g.,][]{2015ApJ...813...88Z, 2016ApJ...816L..12D}, contrary to what is observed for DR Tau. Thus, a stellar companion is also unlikely to be responsible for the arm.  

An infalling stream is a plausible explanation for the molecular arm, given that similar large-scale structures have been detected in association with a number of embedded Class 0/I sources as well as Class II disks proposed to be undergoing late infall \citep[e.g.,][]{2012AA...547A..84T, 2019ApJ...880...69Y, 2020NatAs...4.1158P, 2021ApJS..257...19H, 2022AA...658A.104G, 2022AA...667A..12V}. One possible difference of note is that the structures proposed to be infalling streams in the other systems have tended to be open, whereas DR Tau's pitch angle exhibits a marked decrease with distance from the star. However, this apparent difference may simply be a projection effect, since we do not know their three-dimensional orientations.

As noted in the previous subsection, it is uncertain whether DR Tau is gravitationally unstable. The possibility that DR Tau's arm arises from gravitational instabilities remains intriguing given that clumpy arms are a hallmark of simulations of fragmenting disks \citep[e.g.,][]{2012ApJ...746..110Z, 2012ApJ...750...30B}. Furthermore, migration of clumps onto stars has been proposed as a trigger for FUor outbursts \citep[e.g.,][]{2010Icar..207..509B}. Clump migration might likewise explain DR Tau's extreme brightening event in the 1970s. If DR Tau's disk mass has been estimated correctly, then the presence of a clump along DR Tau's arm raises the question of whether fragmentation can occur under less stringent conditions than models demand. 

Close stellar encounters can also generate large-scale arm-like structures with pitch angles comparable to that observed for the DR Tau molecular arm \citep[e.g.,][]{2015MNRAS.449.1996D, 2019MNRAS.483.4114C, 2020MNRAS.491..504C}. However, \citet{2022ApJS..263...31S} inferred from an analysis of Gaia EDR3 data \citep{2021AA...649A...1G} that the closest expected approach between DR Tau and a neighboring star in the past 10,000 years is $\sim10^5$ au, which would be too distant to meaningfully perturb the known circumstellar environment of DR Tau. \citet{2022AA...658A..63M} found that DQ Tau may have passed within 5100 au of DR Tau 0.23 Myr ago, but considered such an encounter unlikely to be responsible for DR Tau's spiral arms because flyby-induced arms are only expected to survive on timescales of several thousand years \citep[e.g.,][]{2022arXiv220709752C}. 

\subsection{Connections to EXor and FUor phenomena}

In recent years, the circumstellar environments of a number of FUors and EXors have been spatially resolved with millimeter interferometry and high-contrast scattered light imaging. FUors are often associated with envelopes, outflows, and arm-like structures \citep[e.g.,][]{2016SciA....2E0875L, 2017MNRAS.465..834Z, 2017MNRAS.468.3266R, 2017ApJ...843...45K}, similar to the structures associated with DR Tau. In FUor systems, the envelopes supply infalling material that may help to activate gravitational instabilities (and then possibly magnetorotational instabilities), the arms may form as a consequence of gravitational instabilities, and instabilities may trigger outbursts that subsequently drive outflows \citep[e.g.,][]{1994ApJ...424..793E, 2005ApJ...633L.137V, 2010ApJ...713.1143Z}.  With the exception of EX Lup and V1647 Ori, the latter of which is sometimes considered to be an FUor source, the EXor sources imaged so far have generally lacked analogous features \citep[e.g.,][]{2018MNRAS.473..879P, 2018ApJ...859..111H, 2018MNRAS.474.4347C, 2020ApJ...900....7H}. However, DR Tau exhibits striking similarities to EX Lup in that they both feature outflows, non-Keplerian spiral-like structures, and (remnant) envelopes. The circumstellar environments of DR Tau and EX Lup also share similarities with that of RU Lup, which is not classified as an EXor source but is nevertheless an exceptionally active T Tauri star \citep{1945ApJ...102..168J, 1974AA....33..399G, 2020ApJ...898..140H}. \citet{2018ApJ...859..111H} suggested that the presence of complex structures associated with EX Lup but not other EXors is an indication that EX Lup occupies an intermediate evolutionary stage between FUors and most EXors. The same may hold true for DR Tau (and perhaps RU Lup). Alternatively, the differences in EXor circumstellar environments may indicate that EXors are a heterogeneous group of objects, only some of which are closely related to the FUor phenomenon. In any case, the observations of EX Lup, DR Tau, and RU Lup motivate more spatially resolved imaging of extremely active T Tauri stars to elucidate the connection between circumstellar environments and stellar properties. 

\subsection{A changing view of Class II disks}

In the past decade, the introduction of high angular resolution imaging at millimeter wavelengths has transformed our understanding of planet formation by showing that dust substructures on scales of several au are common \citep[e.g.,][]{2015ApJ...808L...3A, 2018ApJ...869L..41A}. Meanwhile, sensitive molecular imaging at more modest resolution has highlighted a deficit in our understanding of disk environments on scales of tens to thousands of au. With single-dish telescopes and earlier generations of interferometers, signs of large-scale non-Keplerian emission towards Class II disks were often ascribed to foreground contamination \citep[e.g.,][]{2001ApJ...561.1074T, 2009ApJ...698..131H, 2011ApJ...734...98O}. Even in the era of more powerful millimeter interferometers, insufficient integration times or insufficient $uv$ coverage at larger spatial scales can lead to key structures being missed. 

High-quality molecular mapping, though, has demonstrated that there are indeed large-scale tails, spirals, streams, and/or remnant envelopes associated with a number of Class II systems \citep[e.g.,][]{2019AJ....157..165A, 2021ApJS..257...19H, 2021ApJ...914...88P, 2022ApJ...930..171H}. Scattered light imaging has also played an important role in uncovering examples of Class II disks that appear to be interacting with surrounding material \citep[e.g.,][]{2004ASPC..321..244G, 2018AA...620A..94G, 2021ApJ...908L..25G}, although as demonstrated by the examples of DR Tau from this work and RU Lup from \citet{2020ApJ...898..140H}, molecular observations can reveal structures far beyond the detected extent of scattered light features. Infall from these larger-scale structures is increasingly being invoked to explain certain disk structures observed at smaller scales, such as misalignments or spiral arms \citep[e.g.,][]{2021ApJ...908L..25G, 2021ApJ...914...88P,2022AA...658A..63M}. These observations thus imply an intriguing link between dynamical processes operating on disparate size scales. For individual systems, though, infall is only one of several possible explanations for the observed disk phenomena. More systematic molecular line observations will be key for establishing patterns of association between large- and small-scale properties.

\section{Summary\label{sec:summary}}
We present new NOEMA observations of $^{12}$CO, $^{13}$CO, C$^{18}$O, SO, DCO$^+$, and H$_2$CO toward the T Tauri star DR Tau, representing the highest-quality millimeter line observations of this source to date. Our findings are as follows:

\begin{enumerate}
    \item CO emission shows that the DR Tau protoplanetary disk is associated with an envelope, a faint asymmetric outflow, and a large non-Keplerian spiral arm with a clump. 
    \item The molecular spiral arm resembles a scaled-up version of the spiral arms detected in scattered light, although the angular resolution of NOEMA is not sufficient to determine whether the molecular arm is an extension of one of the scattered light arms or a separate feature. Whereas the scattered light arms are only detected up to $\sim220$ au in projection from DR Tau, the molecular arm is detected up to $\sim1200$ au in projection from the star. 
    \item We report detections of SO, DCO$^+$, and H$_2$CO in the DR Tau disk for the first time. Their kinematics and compact emission extent suggest that they primarily trace the Keplerian circumstellar disk. 
    \item SO emission is stronger on the northern, redshifted side of the disk. This asymmetry might be linked to infall from an asymmetric envelope or to unresolved spiral substructure associated with the arms detected in scattered light. Higher angular resolution observations of SO will be needed to clarify the origins of the asymmetry. 
\end{enumerate}

DR Tau's envelope, outflow, and arm are reminiscent of the structures that have been observed in association with various FUor sources as well as the EXor source EX Lup. Given that FUor and EXor outbursts have been linked to instabilities driven by envelope accretion, a similar mechanism may account for DR Tau's dramatic stellar brightness changes. The NOEMA observations of DR Tau highlight the utility of sensitive, spatially resolved molecular line observations for providing context about the conditions under which young stars and their protoplanetary disks evolve. 

\begin{acknowledgements}
This work is based on observations carried out under project number W20BE with the IRAM NOEMA Interferometer. IRAM is supported by INSU/CNRS (France), MPG (Germany) and IGN (Spain). This work is also based on observations collected at the European Southern Observatory under ESO programme(s) 0102.C-0453(A). We thank our NOEMA local contact, Ana Lopez-Sepulcre, for setting up the observing scripts and assisting with data reduction. We also thank Arthur Bosman, Ke Zhang, Joel Bregman, Lee Hartmann, Merel van't Hoff, Ardjan Sturm, Melissa McClure, and Ewine van Dishoeck for helpful discussions. We thank the referee, Ruobing Dong, for helpful comments improving the clarity of the manuscript. Support for J. H. was provided by NASA through the NASA Hubble Fellowship grant \#HST-HF2-51460.001-A awarded by the Space Telescope Science Institute, which is operated by the Association of Universities for Research in Astronomy, Inc., for NASA, under contract NAS5-26555. This project has received funding from the European Research Council (ERC) under the European Union's Horizon 2020 research and innovation programme (grant agreement No. 101002188). 
\end{acknowledgements}

\facilities{NOEMA}

\software{\texttt{analysisUtils} (\url{https://casaguides.nrao.edu/index.php/Analysis_Utilities}), \texttt{AstroPy} \citep{2013AA...558A..33A}, \texttt{CASA} \citep{2022PASP..134k4501C}, \texttt{cmasher} \citep{cmasher}, \texttt{emcee} \citep{2013PASP..125..306F}, \texttt{GILDAS} \citep{2005sf2a.conf..721P, 2013ascl.soft05010G}, \texttt{matplotlib} \citep{Hunter:2007}, \texttt{pandas} \citep{the_pandas_development_team_2022_7344967, mckinney-proc-scipy-2010}, \texttt{scikit-image} \citep{scikit-image}, \texttt{SciPy} \citep{2020SciPy-NMeth}}

\appendix

\section{Spectroscopic Parameters of Targeted Lines\label{sec:spectroscopic}}
The spectroscopic parameters of the targeted lines, taken from the Cologne Database for Molecular Spectroscopy \citep{2001AA...370L..49M, 2005JMoSt.742..215M} via Splatalogue\footnote{\url{https://splatalogue.online//}}, are listed in Table \ref{tab:spectroscopic}. Primary line targets are marked in bold. 

\begin{deluxetable*}{ccc}
\tablecaption{Spectroscopic Parameters of All Targeted Lines\label{tab:spectroscopic}}
\tablehead{
\colhead{Transition}&\colhead{Rest frequency}&\colhead{$E_u$}\\
&(GHz)&(K)}
\startdata
$^{13}$C$^{17}$O $J=2-1$&214.5738730 & 15.4 \\
\textbf{SO $J_N = 5_5-4_4$}& 215.2206530 & 44.1 \\
\textbf{DCO$^+$ $J=3-2$} &216.1125822 & 20.7 \\
H$_2$S $J_{K_aK_c} = 2_{20}-2_{11}$& 216.7104365 & 84.0\\
$c$-C$_3$H$_2$ $J_{K_aK_c}=3_{30}-2_{21}$  & 216.2787560 & 19.5\\
SiO $J=5-4$ & 217.1049190& 31.3\\
DCN $J=3-2$ & 217.2385378 & 20.9 \\
$c$-C$_3$H$_2$ $J_{K_aK_c}=5_{14}-4_{23}$ &217.9400460 & 35.4\\
\textbf{H$_2$CO $J_{K_aK_c}=3_{03}-2_{02}$} & 218.2221920& 21.0\\
HC$_3$N $J=24-23$ & 218.3247230&131.0\\ 
\textbf{H$_2$CO $J_{K_aK_c}=3_{22}-2_{21}$} & 218.4756320 & 68.1\\
\textbf{H$_2$CO $J_{K_aK_c}=3_{21}-2_{20}$} & 218.7600660&68.1\\
\textbf{C$^{18}$O $J=2-1$} &219.5603541& 15.8\\
\textbf{SO $J_N = 6_5-5_4$}& 219.9494420 & 35.0\\
\textbf{$^{13}$CO $J=2-1$} &220.3986842 &15.9\\
\textbf{$^{12}$CO $J=2-1$} &230.5380000&16.6\\
OCS $J=19-18$ & 231.0609934 &110.9\\
N$_2$D$^+$ $J=3-2$ & 231.3218283&22.2\\
$^{13}$CS $J=5-4$ & 231.2206852&33.3\\
C$_2$S $J_N=19_{18}-18_{17}$ & 233.9384580 & 109.6 \\
PN $J= 5-4$& 234.9356940 &33.8\\
HC$_3$N $J=26-25$ & 236.5127888 & 153.2\\
H$_2$CS $J_{K_aK_c}=7_{17}-6_{16}$ & 236.7270204&58.6\\
\enddata
\end{deluxetable*}

\section{Channel Maps\label{sec:chanmaps}}
\begin{figure*}
\figurenum{10.1}
\begin{center}
\includegraphics{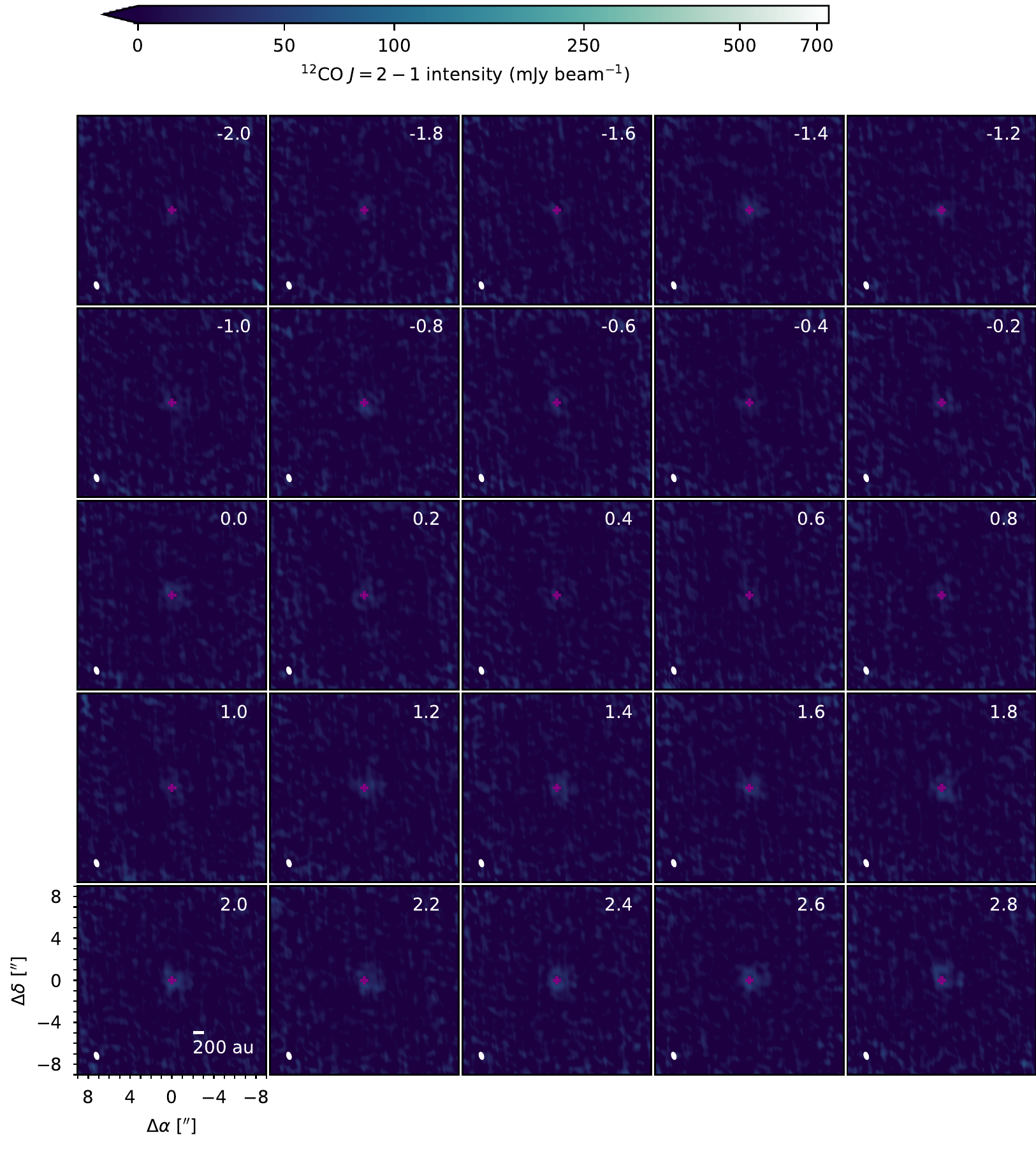}
\end{center}
\caption{Channel maps of $^{12}$CO $J=2-1$ toward DR Tau, part 1. The top right of each panel is labelled with the LSRK velocity (km s$^{-1}$). The synthesized beam is drawn in the lower left corner of each panel. The purple crosses denote the disk center. Offsets from the disk center (in arcseconds) are marked on the axes in the lower left corner. The color scale uses an arcsinh stretch to make faint extended features more visible.\label{fig:chanmaps}}
\end{figure*}

Channel maps of the primary line targets (listed in Table \ref{tab:imageproperties}) are presented in Figure \ref{fig:chanmaps}.

\begin{figure*}
\figurenum{10.1}
\begin{center}
\includegraphics{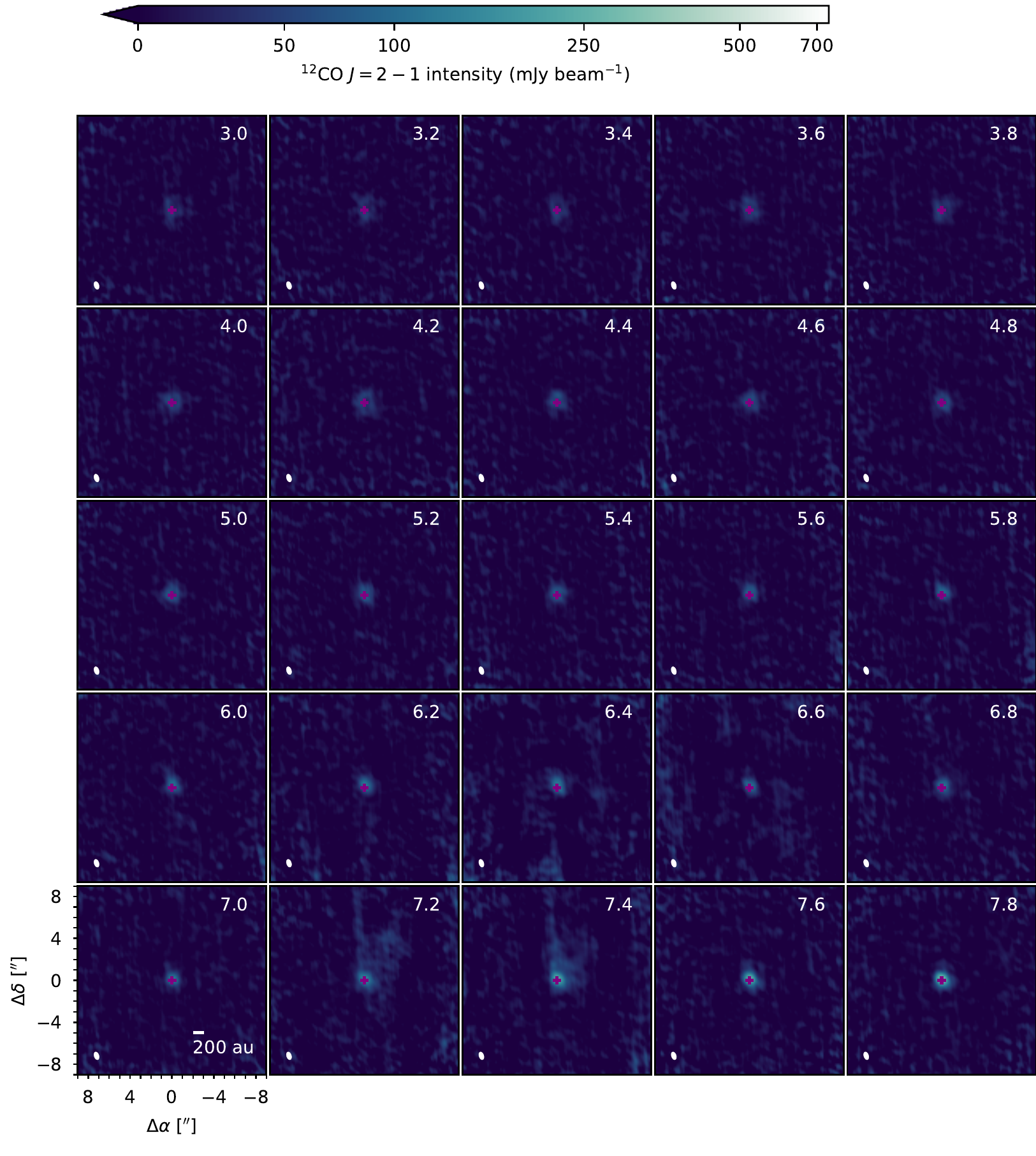}
\end{center}
\caption{Channel maps of $^{12}$CO $J=2-1$ toward DR Tau, part 2. }
\end{figure*}

\begin{figure*}
\figurenum{10.1}
\begin{center}
\includegraphics{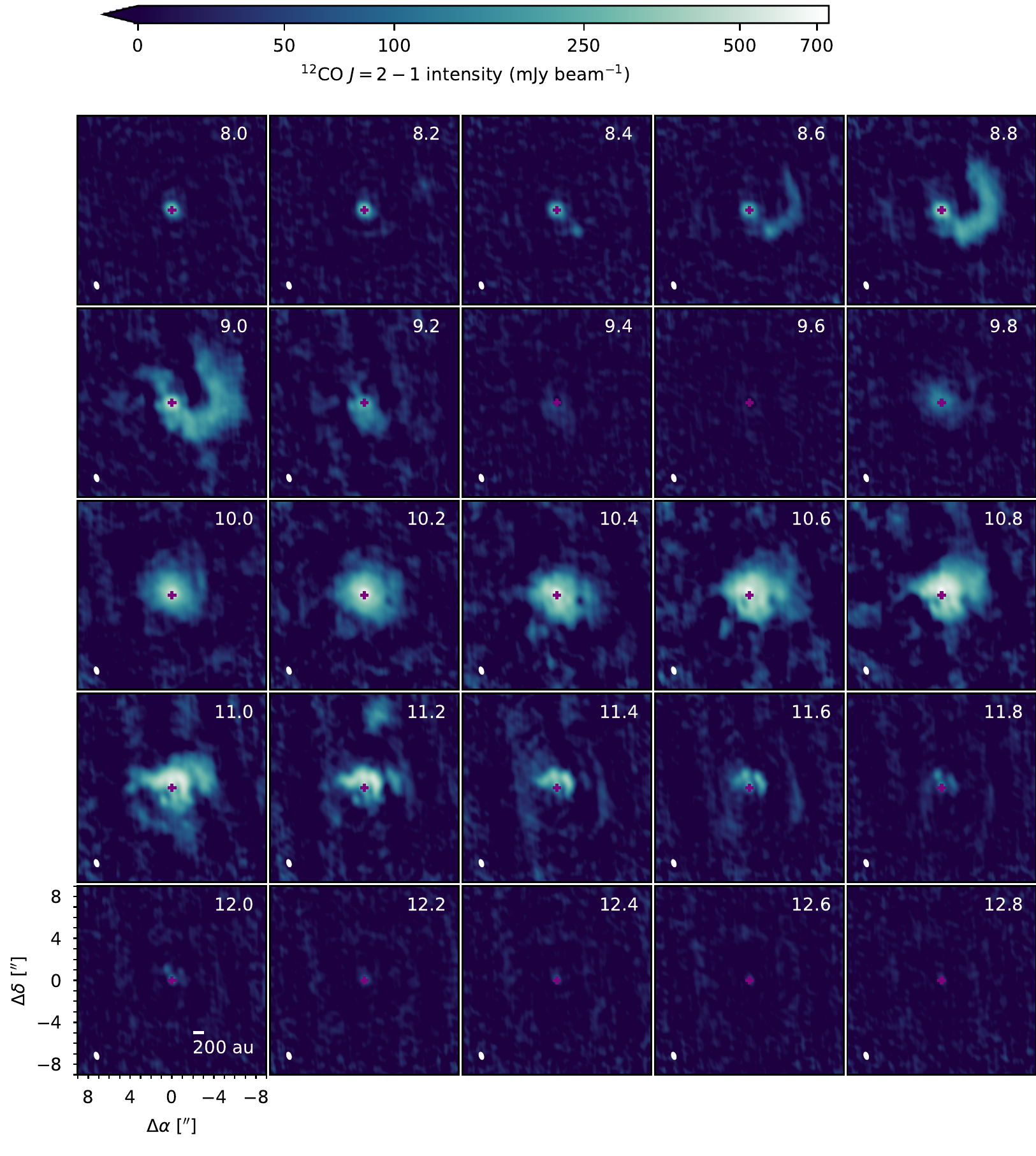}
\end{center}
\caption{Channel maps of $^{12}$CO $J=2-1$ toward DR Tau, part 3. }
\end{figure*}

\begin{figure*}
\figurenum{10.1}
\begin{center}
\includegraphics{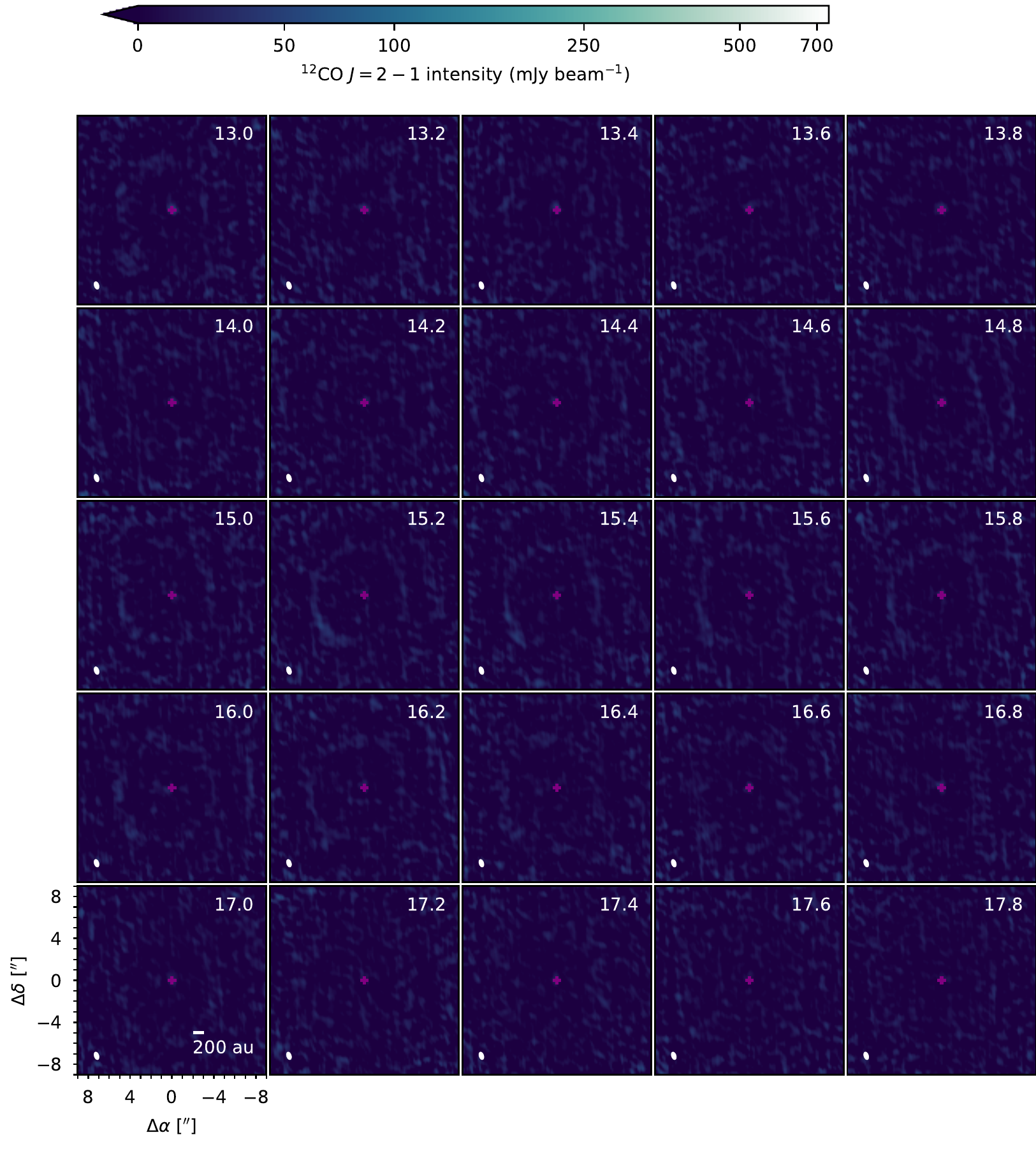}
\end{center}
\caption{Channel maps of $^{12}$CO $J=2-1$ toward DR Tau, part 4. }
\end{figure*}

\begin{figure*}
\figurenum{10.2}
\begin{center}
\includegraphics{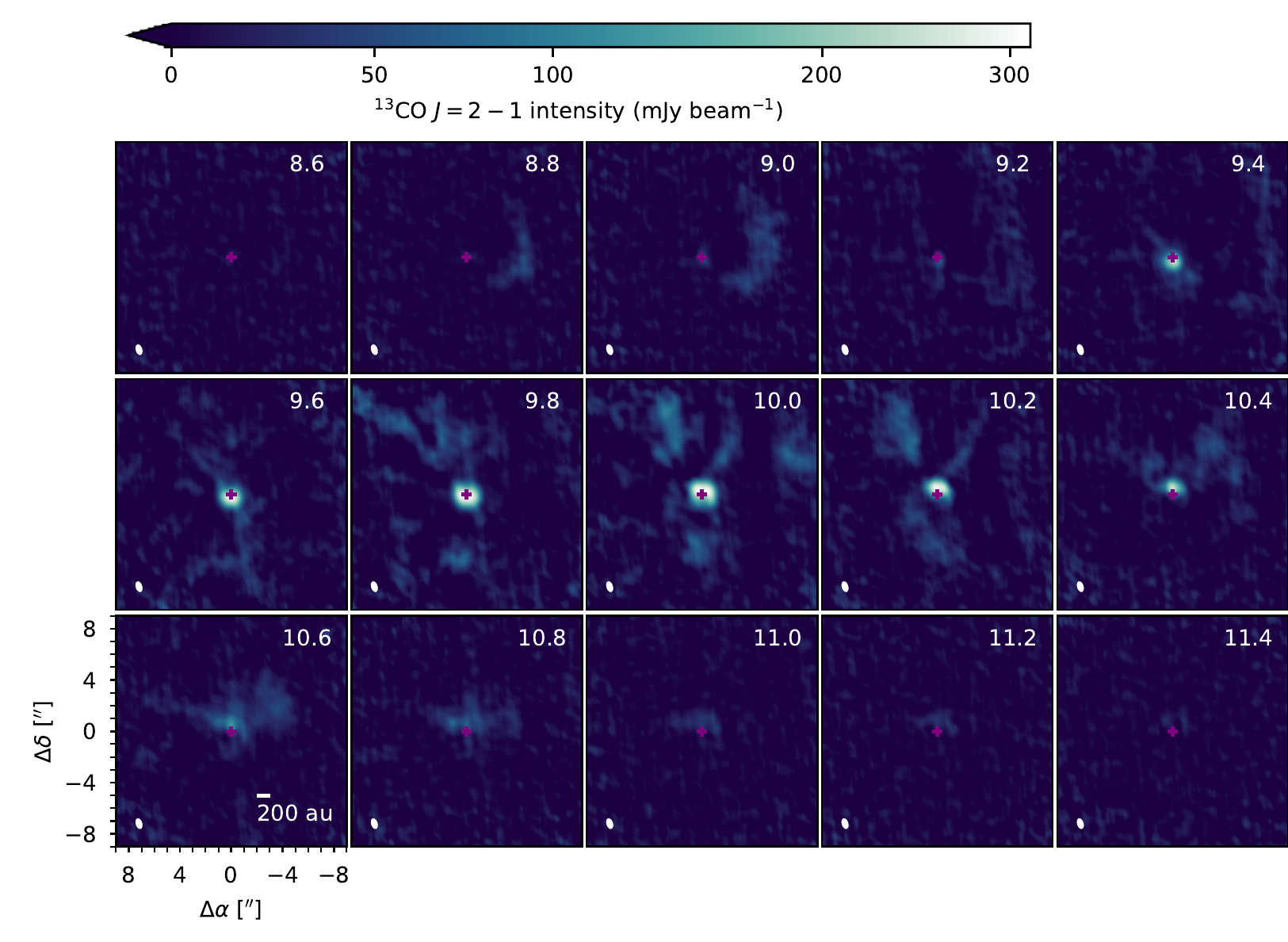}
\end{center}
\caption{Channel maps of $^{13}$CO $J=2-1$ toward DR Tau. }
\end{figure*}

\begin{figure*}
\figurenum{10.3}
\begin{center}
\includegraphics{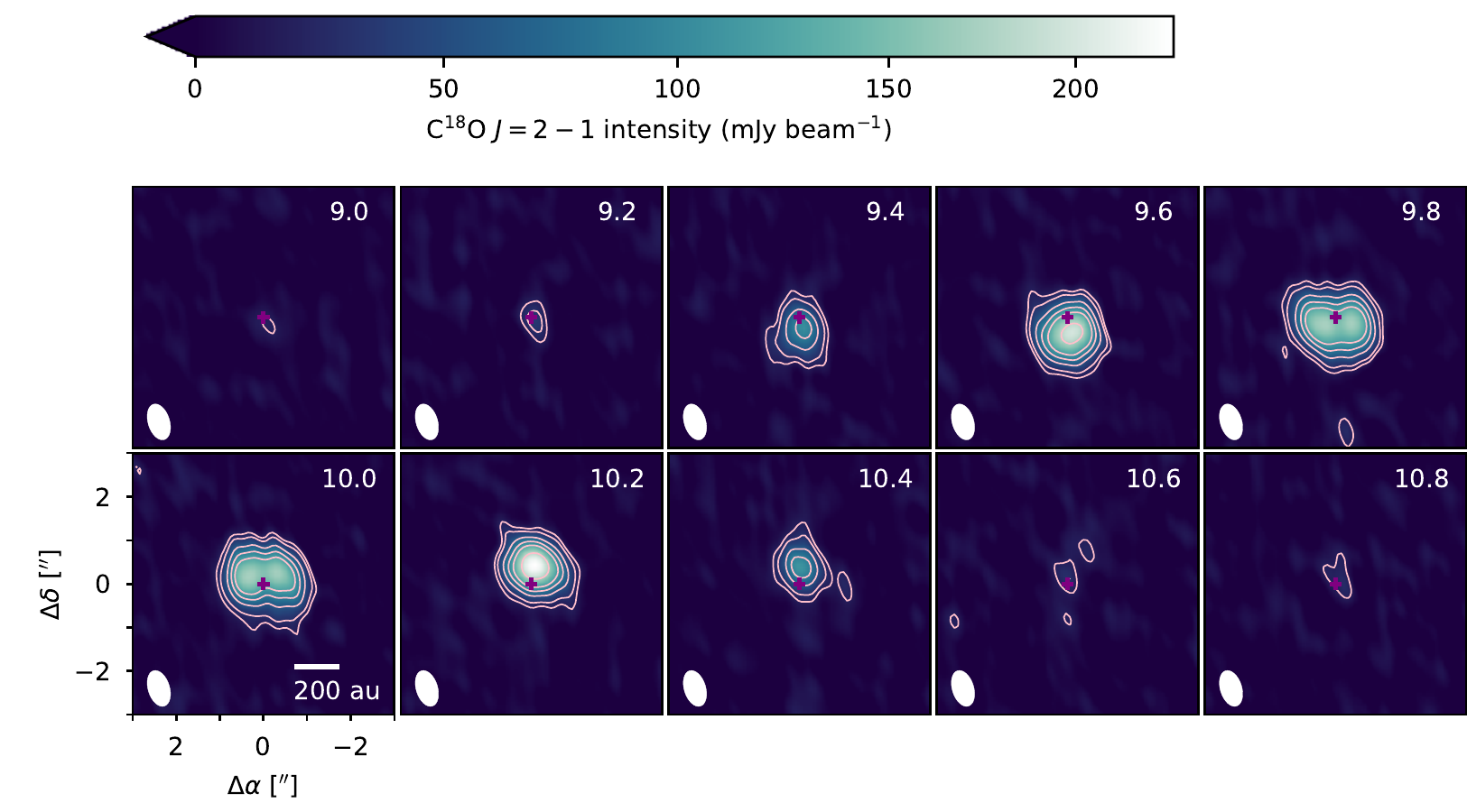}
\end{center}
\caption{Channel maps of C$^{18}$O $J=2-1$ toward DR Tau. Contours are drawn in pink at the $3,5,10,15,20, 30\sigma$ levels.}
\end{figure*}

\begin{figure*}
\figurenum{10.4}
\begin{center}
\includegraphics{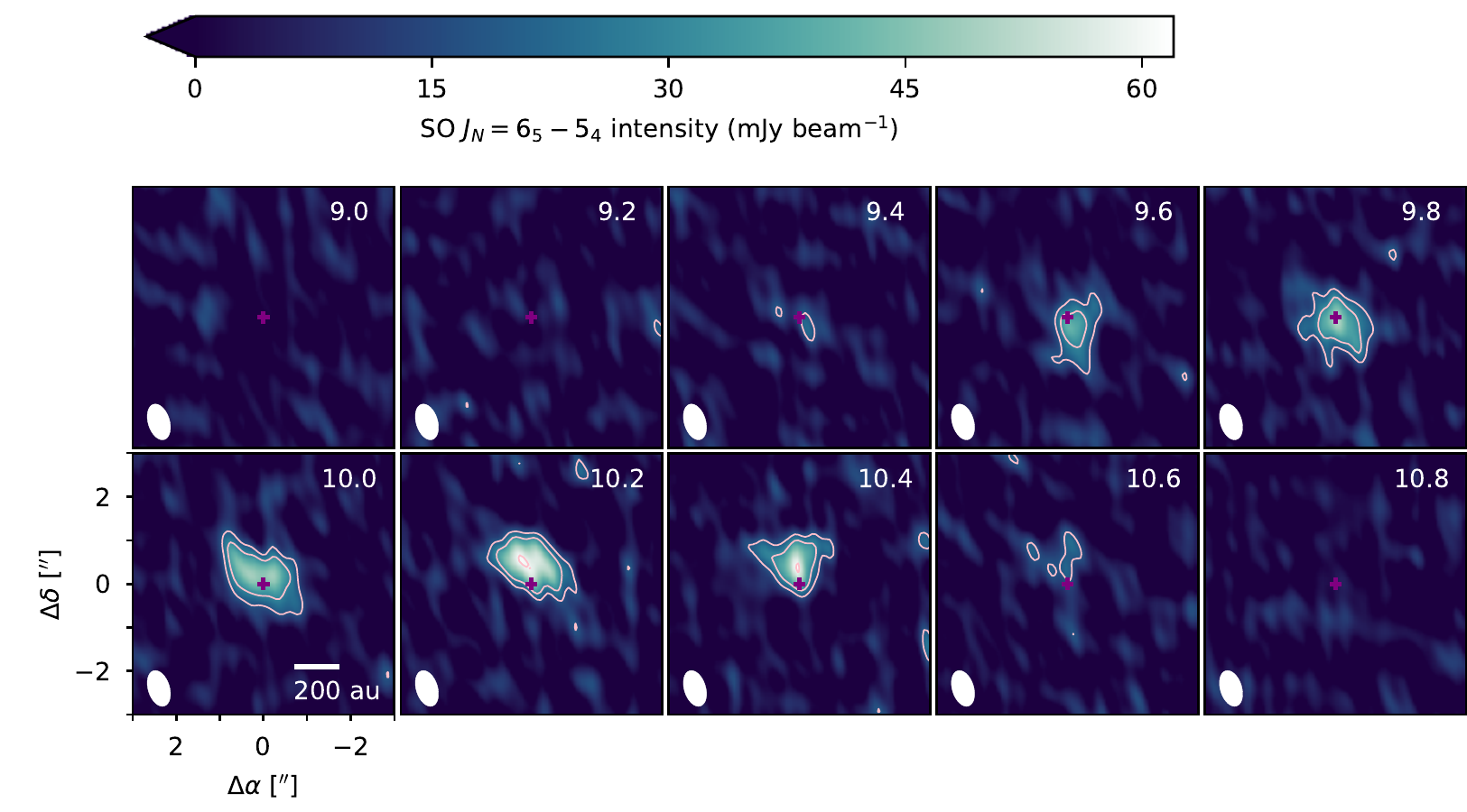}
\end{center}
\caption{Channel maps of SO $J_N=6_5-5_4$ toward DR Tau. Contours are drawn in pink at the $3,5, 10\sigma$ levels.}
\end{figure*}

\begin{figure*}
\figurenum{10.5}
\begin{center}
\includegraphics{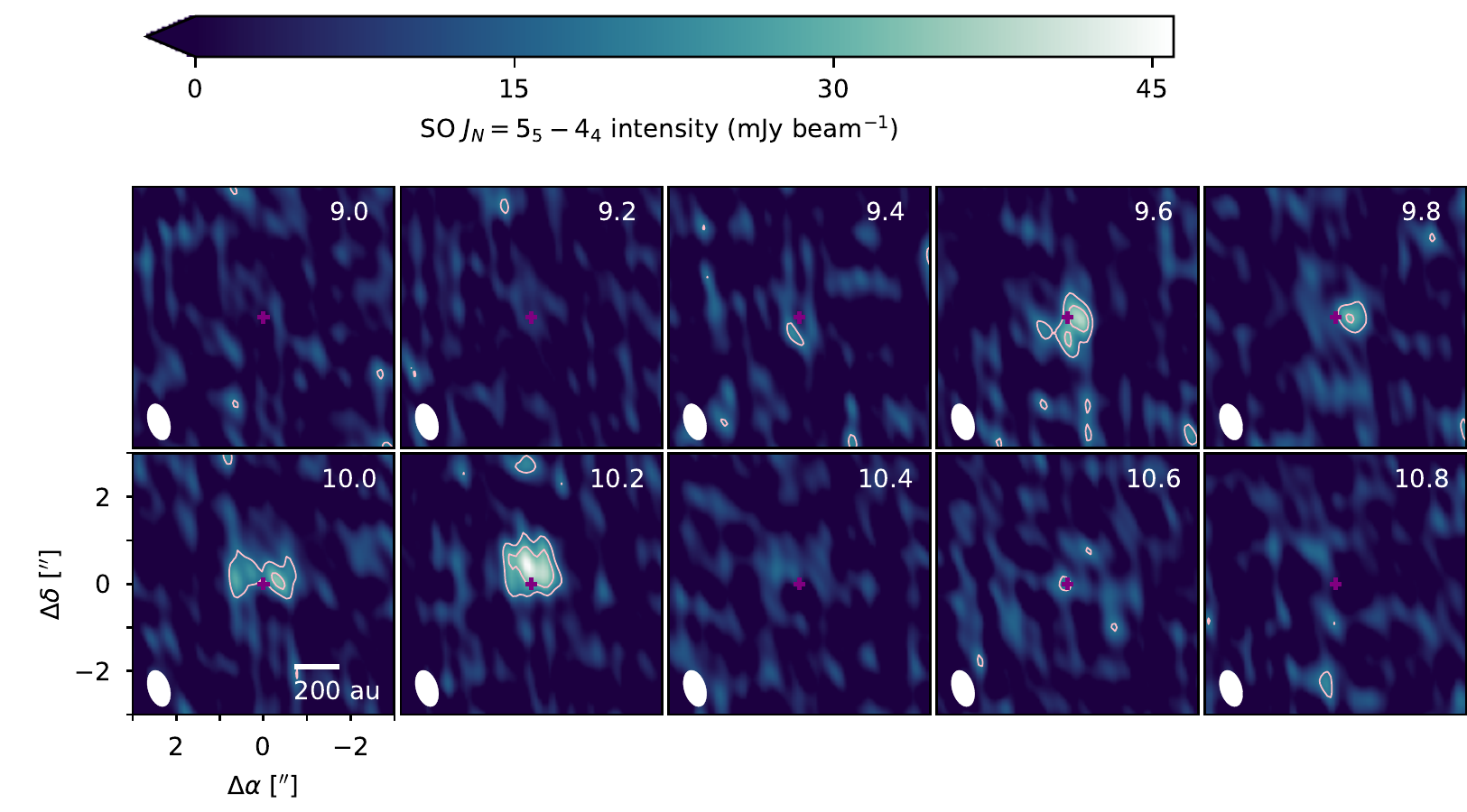}
\end{center}
\caption{Channel maps of SO $J_N=5_5-4_4$ toward DR Tau. Contours are drawn in pink at the $3,5\sigma$ levels.}
\end{figure*}

\begin{figure*}
\figurenum{10.6}
\begin{center}
\includegraphics{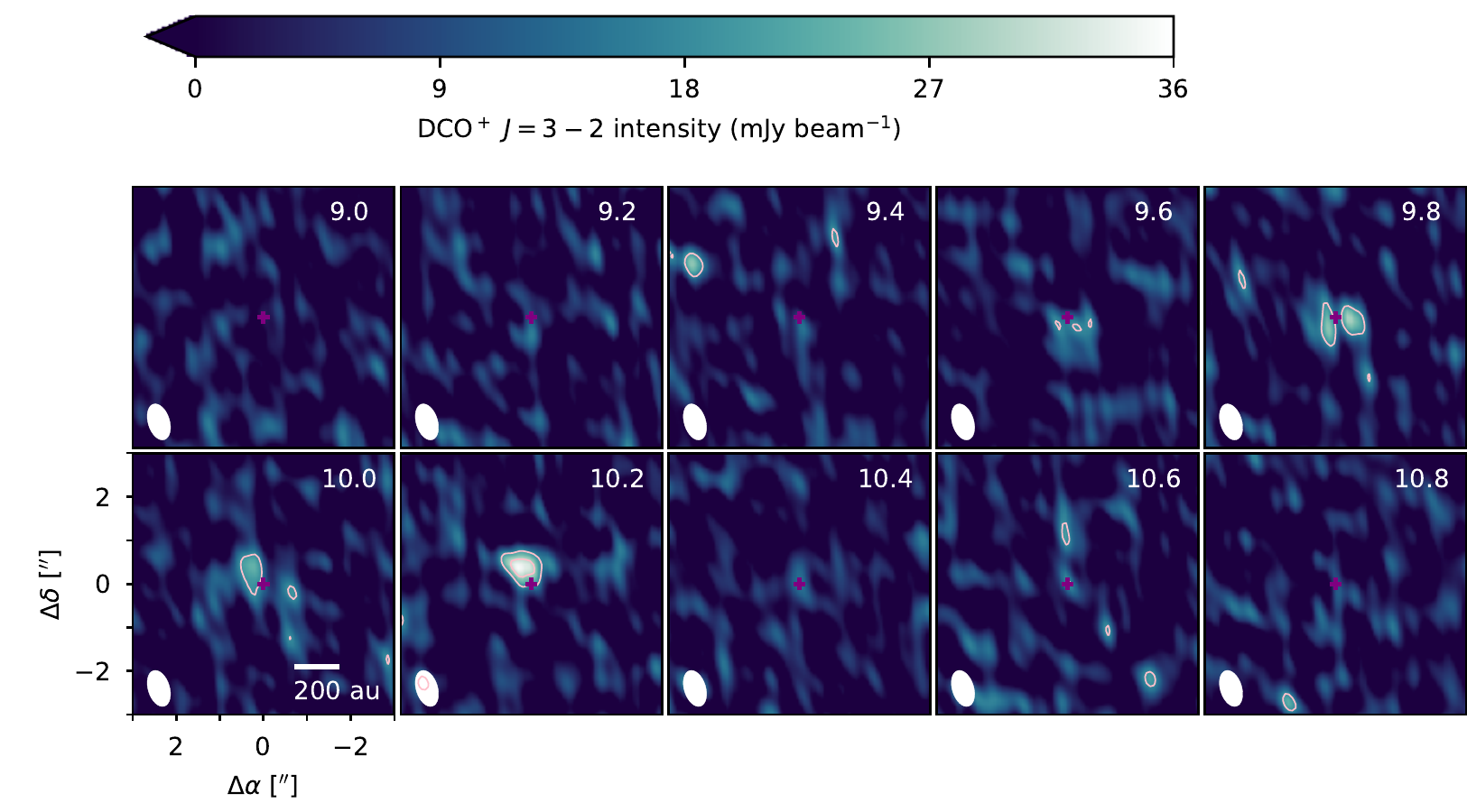}
\end{center}
\caption{Channel maps of DCO$^+$ $J=3-2$ toward DR Tau. Contours are drawn in pink at the $3,5\sigma$ levels.}
\end{figure*}

\begin{figure*}
\figurenum{10.7}
\begin{center}
\includegraphics{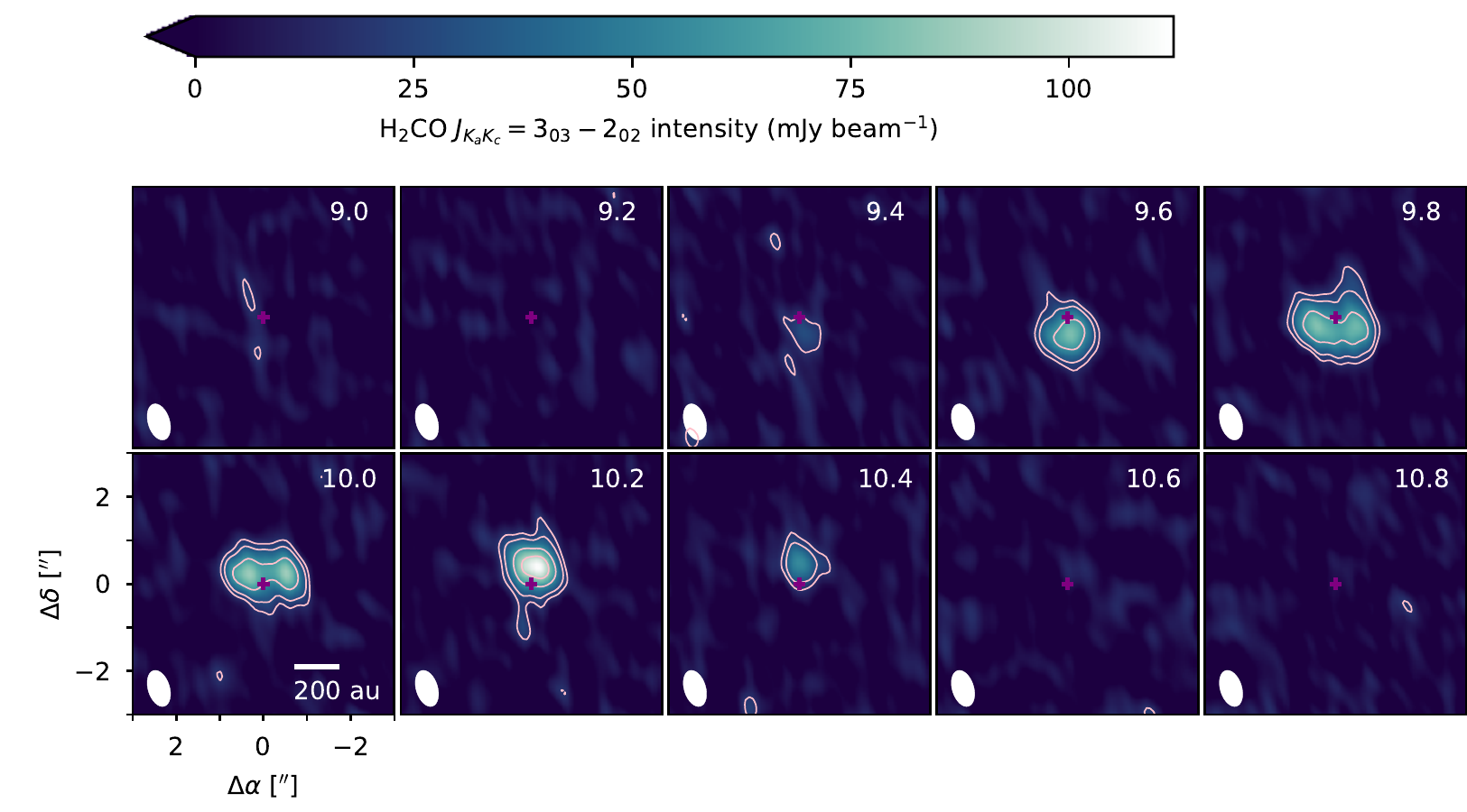}
\end{center}
\caption{Channel maps of H$_2$CO $J_{K_aK_c}=3_{03}-2_{02}$ toward DR Tau. Contours are drawn in pink at the $3,5,10,15\sigma$ levels.}
\end{figure*}

\begin{figure*}
\figurenum{10.8}
\begin{center}
\includegraphics{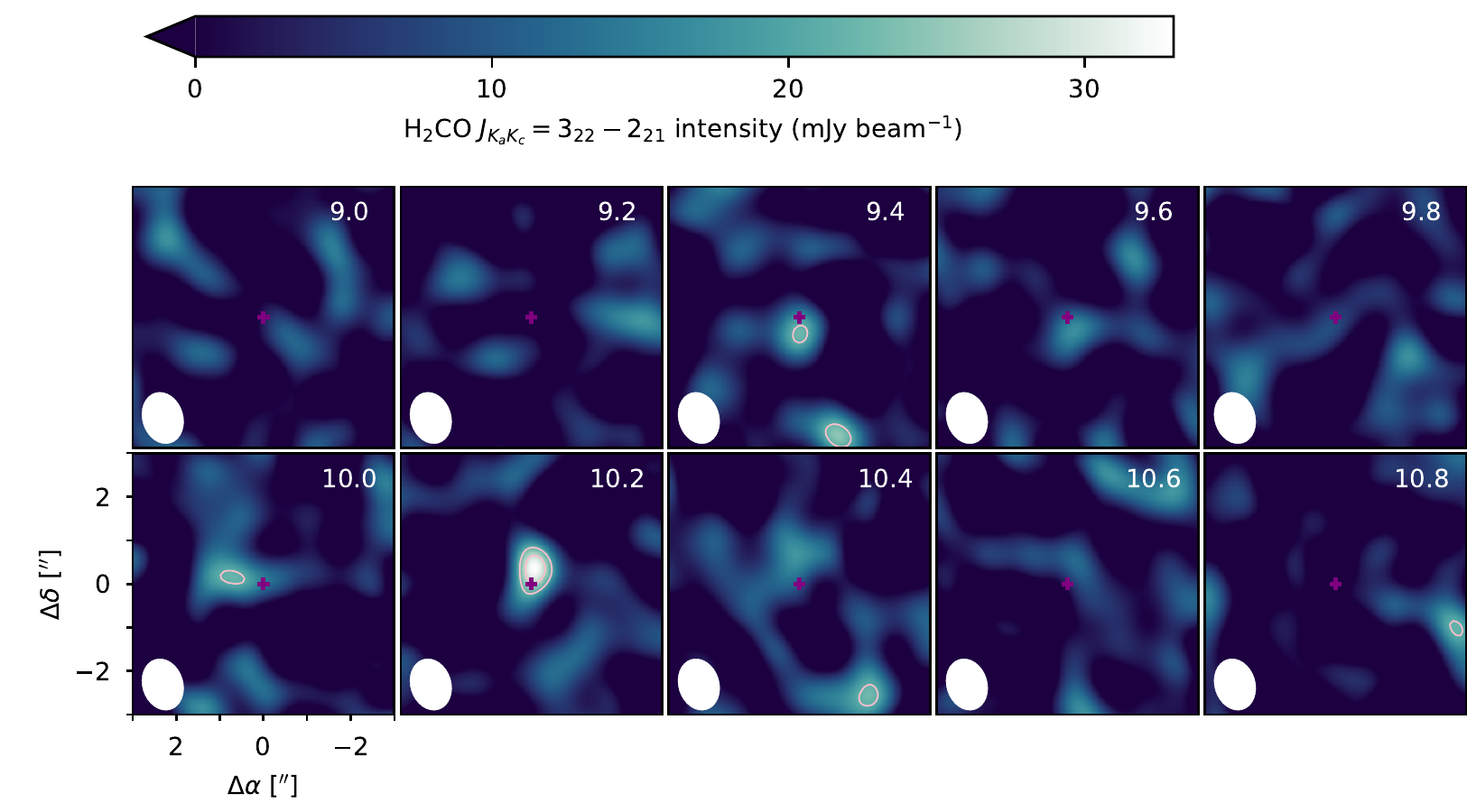}
\end{center}
\caption{Channel maps of H$_2$CO $J_{K_aK_c}=3_{22}-2_{21}$ toward DR Tau. Contours are drawn in pink at the $3,4\sigma$ levels.}
\end{figure*}

\begin{figure*}
\figurenum{10.9}
\begin{center}
\includegraphics{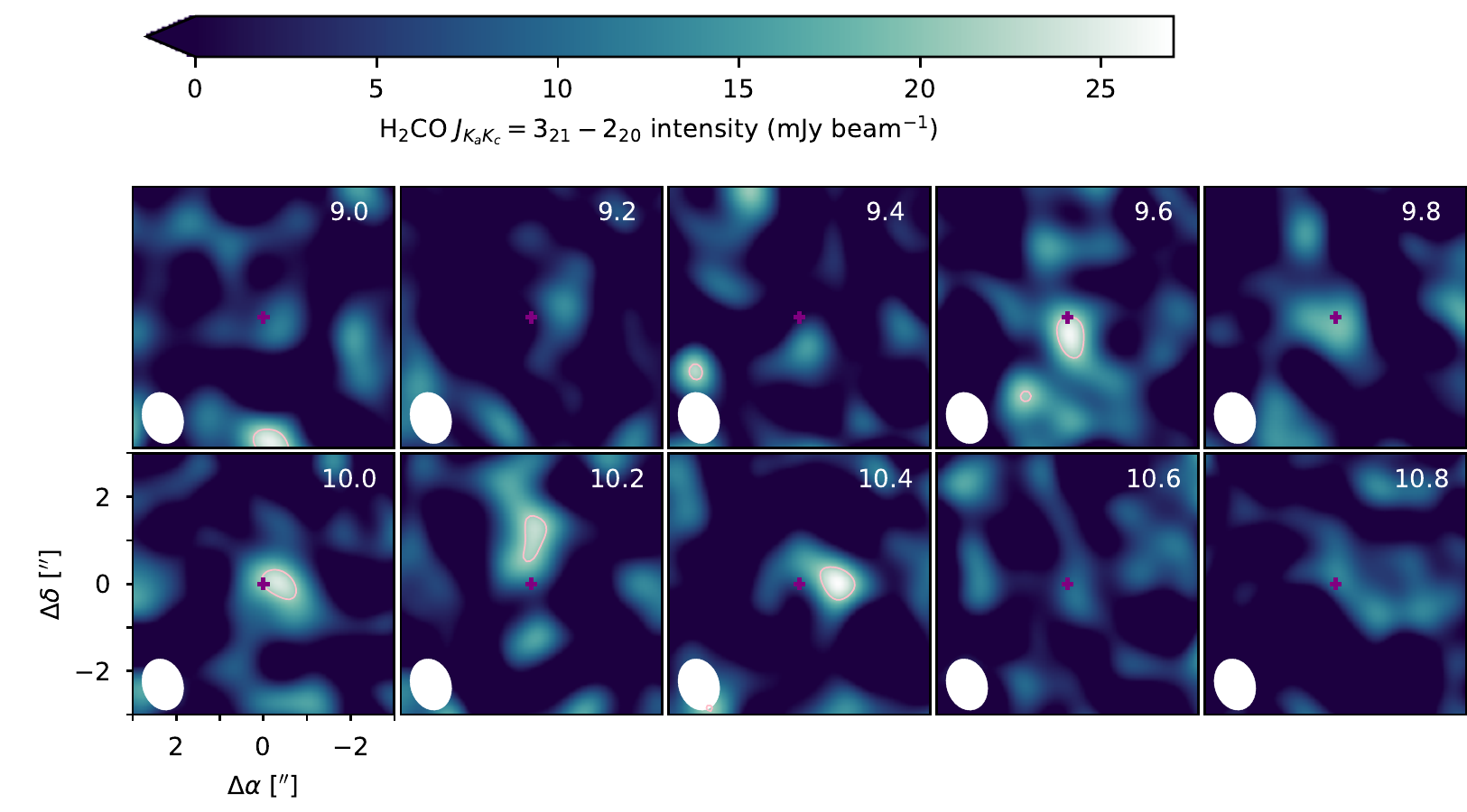}
\end{center}
\caption{Channel maps of H$_2$CO $J_{K_aK_c}=3_{21}-2_{20}$ toward DR Tau. Contours are drawn in pink at the $3\sigma$ level.}
\end{figure*}

 \section{Auxiliary line targets\label{sec:auxiliarylines}}
 Table \ref{tab:auxiliaryimageproperties} lists the beam sizes, per-channel rms of the image cubes, and $3\sigma$ flux upper limits for the auxiliary line targets. The flux upper limits were estimated assuming the same velocity range and aperture used to measure the C$^{18}$O flux (see Table \ref{tab:imageproperties}). Flux upper limits may be underestimated if the molecule is primarily present in the envelope or outflow rather than the disk. 

\begin{deluxetable*}{ccccccc}
\tablecaption{Imaging Summary for Auxiliary Line Targets \label{tab:auxiliaryimageproperties}}
\tablehead{
\colhead{Transition}&\colhead{Synthesized beam}&\colhead{Per-channel RMS noise\tablenotemark{a}}&\colhead{$3\sigma$ Flux Upper Limit}\\
&(arcsec $\times$ arcsec ($^\circ$))&(mJy beam$^{-1}$)&(mJy km s$^{-1}$)}
\startdata
$^{13}$C$^{17}$O $J=2-1$ & 1.21 $\times$ 0.93  (18.0$^\circ$)&8 & $<40$\\
H$_2$S  $J_{K_aK_c}=2_{20}-2_{11}$& 1.21 $\times$ 0.93  (16.6$^\circ$)& 7 & $<40$\\
$c$-C$_3$H$_2$ $J_{K_aK_c}=3_{30}-2_{21}$ & 1.21 $\times$ 0.93  (16.7$^\circ$)& 7 & $<30$\\
SiO $J=5-4$& 1.21 $\times$ 0.93  (16.4$^\circ$)&7 & $<30$\\
 DCN $J=3-2$&  1.21 $\times$ 0.93  (16.5$^\circ$)&7 & $<40$ \\
 $c$-C$_3$H$_2$ $J_{K_aK_c}=5_{14}-4_{23}$ &1.21 $\times$ 0.93  (16.7$^\circ$)& 7 & $<30$\\
 HC$_3$N $J=24-23$ & 1.20 $\times$ 0.94  (17.0$^\circ$)& 7& $<30$\\
 OCS $J=19-18$&1.18 $\times$ 0.91  (17.3$^\circ$) & 7 & $<30$ \\
N$_2$D$^+$ $J=3-2$& 1.18 $\times$ 0.91  (17.3$^\circ$) & 10 & $<70$ \\
$^{13}$CS $J=5-4$&1.18 $\times$ 0.91  (17.3$^\circ$) &8 &$<50$ \\
C$_2$S $J_N=19_{18}-18_{17}$  &1.17 $\times$ 0.90  (16.6$^\circ$)& 10 &$<50$\\
PN $J=5-4$&1.17 $\times$ 0.90  (16.8$^\circ$) & 9 & $<40$\\
HC$_3$N $J=26-25$ & 1.17 $\times$ 0.90  (16.7$^\circ$)& 10 & $<50$\\
H$_2$CS  $J_{K_aK_c}=7_{17}-6_{16}$&1.17 $\times$ 0.90  (16.6$^\circ$)& 11 & $<80$
\enddata
\tablenotetext{a}{For channel widths of 0.2 km s$^{-1}$.}
\end{deluxetable*}

\end{CJK*}
\end{document}